\documentclass[aps,prx,twocolumn,showpacs,showkeys,superscriptaddress,reprint]{revtex4-1} 
\usepackage{graphics} 
\usepackage{graphicx} 
\usepackage{amsmath} 
\usepackage{color}
\usepackage{lineno}
\usepackage{epstopdf}

\begin{document} 

\title{Giant rectification in strongly-interacting driven tilted systems} 
\author{Juan Jos\'e Mendoza-Arenas}
\email{jum151@pitt.edu} 
\affiliation{Department of Mechanical Engineering and Materials Science, University of Pittsburgh, Pittsburgh, Pennsylvania 15261, USA}
\affiliation{Department of Physics and Astronomy, University of Pittsburgh, Pittsburgh, Pennsylvania 15261, USA}
\affiliation{H. H. Wills Physics Laboratory, University of Bristol, Bristol, BS8 1TL, United Kingdom}
\author{Stephen R. Clark} 
\email{stephen.clark@bristol.ac.uk} 
\affiliation{H. H. Wills Physics Laboratory, University of Bristol, Bristol, BS8 1TL, United Kingdom}

\date{\today} 

\begin{abstract}
Correlated quantum systems feature a wide range of nontrivial effects emerging from interactions between their constituting particles. In nonequilibrium scenarios, these manifest in phenomena such as many-body insulating states and anomalous scaling laws of currents of conserved quantities, crucial for applications in quantum circuit technologies. In this work we propose a giant rectification scheme based on the asymmetric interplay between strong particle interactions and a tilted potential, each of which induces an insulating state on their own. While for reverse bias both cooperate and induce a strengthened insulator with an exponentially suppressed current, for forward bias they compete generating conduction resonances; this leads to a rectification coefficient of many orders of magnitude. We uncover the mechanism underlying these resonances as enhanced coherences between energy eigenstates occurring at avoided crossings in the system's bulk energy spectrum. Furthermore, we demonstrate the complexity of the many-body nonequilibrium conducting state through the emergence of enhanced density matrix impurity and operator space entanglement entropy close to the resonances. Our proposal paves the way for implementing a perfect diode in currently-available electronic and quantum simulation platforms.
\end{abstract} 

\maketitle 

\section{Introduction}
Due to their vast and exciting phenomenology, as well as the rapid advances on their control protocols, many-body quantum systems have become key ingredients for the development of novel technologies at the nanoscale. In particular, in nonequilibrium regimes such systems feature properties which make them highly appealing for applications in quantum circuits \cite{benenti2017physrep,bertini2021rmp,pekola2021rmp,landi2021arxiv}. This has been exemplified in several platforms where tunable transport of particles or heat can be induced and characterized, including setups based on electronic devices such as molecular electronics \cite{xin2019nat} and quantum dots \cite{josefsson2018nat,dutta2020prl}, or in quantum simulators such as cold atoms \cite{hausler2021prx,amico2022arxiv} and superconducting qubits \cite{ronzani2018nat,maillet2020nat}. In parallel to these seminal experimental advances, there have been numerous recent theoretical proposals to use quantum systems of interacting particles as different types of circuit elements that efficiently perform specific tasks. This includes autonomous quantum thermal machines \cite{nosotros2020prx,khandelwal2021prx,BohrBrask2022operational}, transistors \cite{marchukov2016nat,wilsmann2018nat}, magnetoresistors \cite{poulsen2021prl}, and a quantum analogue of the Wheatstone bridge \cite{poulsen2022prl}. 

Given their broad applicability, much attention in this field has been directed towards quantum diodes. For such devices, spatial asymmetries and non-linearities are engineered together to allow transport in one direction under a chemical potential or thermal bias, and to suppress it in the opposite direction when the bias is inverted. Experimentally, commercial electronic semiconductor diodes have demonstrated a ratio of forward and backward currents (or rectification coefficient) of $\approx10^5-10^8$ \cite{evers2020rmp}. Meanwhile alternative technologies remain under development including molecular diodes, with coefficients of up to $O(10^5)$ \cite{chenx2017nat}, and those based on superconducting components \cite{yuan2021pnas,lin2022nat,pal2022nat}. Theoretically, several studies have recently proposed different rectification schemes based on quantum systems \cite{pepino2009prl,balachandran2019prl,pereira2019pre,balachandran2019pre,yamamoto2020prr,chioquetta2021pre,upadhyay2021pre,haack2023prr}. These efforts include the setups potentially displaying giant rectification \cite{balachandran2018prl,lee2020entropy,lee2021pre,lee2022pre,poulsen2022pra}, which rely on complicated geometries or potential landscapes to induce rectification coefficients of several orders of magnitude using finely-tuned parameters. 

Here we put forward a rectification scheme which naturally emerges over a broad range of parameters in tilted interacting quantum lattices. These systems are the object of intense research as they have been argued to feature disorder-free (Stark) many-body localization (MBL) \cite{schulz2019prl,nieuwenburg2910pnas,taylor2020prb,yao2021prb,zisling2022prb}, quantum scars \cite{khemani2020prb,desaules2021prl}, and counter-intuitive phenomena such as time crystals \cite{kshetrimayum2020prb} and transport opposite to an applied electric field \cite{klockner2020prl}. Moreover, they have been implemented experimentally in several quantum simulation platforms \cite{morong2021nat,guardado2020prx,scherg2021nat,hebbe2021prx,guo2021prl,kohlert2023prl}, making them attainable for state-of-the-art applications in quantum technologies. Neither the spatial asymmetry of the tilted onsite-potential nor the non-linearity of inter-particle interactions can on their own induce rectification. However, the ability to simultaneously engineer both ingredients in such platforms opens the possibility of its implementation. Here we reveal the underlying mechanism of this interaction-tilt interplay and show how the vastly differing transport properties \cite{doggen2021prb} with the direction of bias give rise to giant rectification. Specifically, we find that it can exceed coefficients of $10^9$ in small systems with no need of fine tuning and even with moderate interactions. We emphasize that in spite of the simplicity of our scheme, it features a rich phenomenology resulting in a genuine many-body rectification response, while also establishing a framework to build more complex nonequilibrium physics on top of it.

\begin{figure*}[t]
\begin{center}
\includegraphics[scale=0.45]{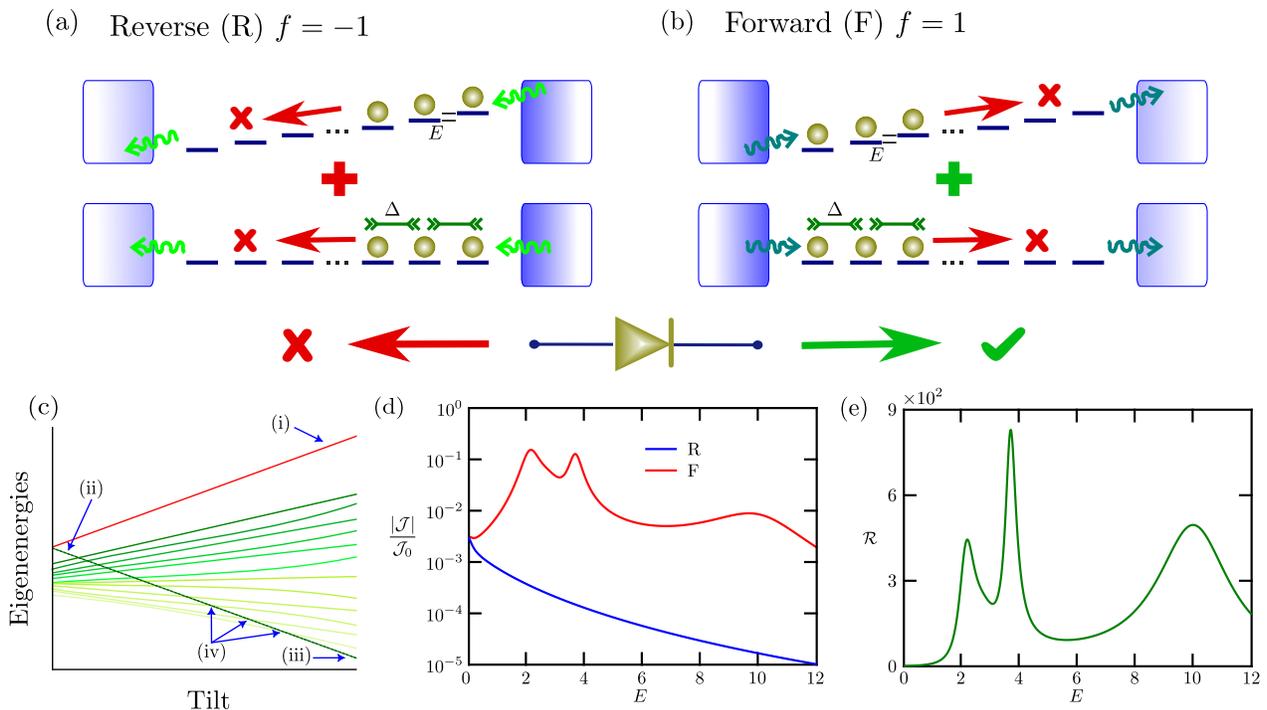}
\end{center}
\caption{Giant rectification in tilted strongly-interacting lattice. (a) For reverse transport, particles are mostly injected from a reservoir on the right boundary of the system, and mostly ejected to a reservoir on the left. For interactions or tilt alone, the resulting state is of high-energy, due to the adjacency of particles in the former case, or due to the particles filling the top of the potential in the latter. An enhanced insulator is induced when both coexist. (b) For forward transport, particles flow from the left to the right reservoir. The tilted noninteracting state is of low energy as particles now fill the bottom of the potential. Thus tilt now competes with interactions, which favors conduction. (c) Sketch of the eigenstructure of a tilted interacting system. \textbf{i}. High-energy insulating eigenstate preferentially populated by reverse bias.
\textbf{ii}. High-energy insulating eigenstate preferentially populated by forward bias and dominant interactions. \textbf{iii}. Low-energy insulating eigenstate preferentially populated by forward bias and very large tilt. \textbf{iv}. The path from the interaction- to the tilt-dominated insulator, indicated by the dashed line, features avoided crossings at intermediate regimes where occupation probability is transferred between eigenstates, enhancing the transport. (d) Forward (F) and reverse (R) particle currents in a lattice of $N=4$ and $\Delta=5$ as a function of the tilt. Currents have been rescaled by the maximum current $\mathcal{J}_0$ obtained with the same driving for a noninteracting homogeneous system (see Eq. \eqref{curr_noE}). (e) The corresponding rectification coefficient $\mathcal{R}$.} 
\label{setup}
\end{figure*}

This article is organized as follows. In Sec. \ref{sec_scheme} we describe the model of our quantum diode, discuss how the transport properties are calculated and illustrate the emergence of large rectification in a small system. In Sec. \ref{sec_insulating} we demonstrate the insulating nature of the system for reverse transport. We explain the mechanism underlying current resonances for forward transport in Sec. \ref{sec_resonances}. The combination of both results leading to giant rectification is discussed in Sec. \ref{sec_recti}. Finally, in Sec. \ref{sec_conclu} we present the conclusions and outlook of our work. Technical details of our calculations are described in the Appendices.

\section{Giant rectification scheme} \label{sec_scheme}
Our quantum diode is based on a boundary-driven configuration, which can be biased from right to left ({\em reverse driving}) or left to right ({\em forward driving}). For reverse driving, sketched in Fig. \ref{setup}(a), both interactions and tilt each support on their own the formation of a high-energy current-blocking particle domain at the right boundary. 
When interactions and tilt coexist they thus cooperate to induce a high-energy state with an enhanced insulating nature. For forward driving, depicted in Fig. \ref{setup}(b), both strong interactions and tilt favour on their own the formation of an insulating domain at the left boundary. For interactions alone this state remains high-energy, while for the tilt alone it is low-energy (see Fig. \ref{setup}(c)). 
By being located at opposite ends of the energy spectrum, interactions and tilt now conflict with each other when they coexist. When both are of similar order, their competition breaks the domains and thus allows particle conduction. This forward-reverse transport asymmetry makes the system behave as a quantum diode. 

\subsection{Boundary-driven model}
To unveil this phenomenon we consider a simple generic model of interacting particles: a one-dimensional lattice of spinless fermions in a tilted potential. The Hamiltonian for $N$ sites is
\begin{align} \label{hami_tilted}
H&=\sum_{j=1}^{N-1}\left[\frac{J}{2}\left(c_j^{\dagger}c_{j+1}+\text{H.c.}\right)+\Delta\left(n_j-\frac{1}{2}\right)\left(n_{j+1}-\frac{1}{2}\right)\right]\notag \\
&\qquad \qquad \qquad \qquad +\sum_{j=1}^N\left(\mu+\frac{E}{2}j\right)\left(n_j-\frac{1}{2}\right),
\end{align}
where $c_j^{\dagger}$ ($c_j$) creates (annihilates) a fermion on site $j$, $n_j=c_j^{\dagger}c_j$ is the corresponding particle number operator, $J$ is the hopping amplitude (we take $J=1$ to set the energy scale), $\Delta$ is the nearest-neighbor density-density interaction, and $\mu$ is the diode's chemical potential. We take the tilt strength $E>0$ so the potential always increases from left to right. For the noninteracting case, this model shows Wannier-Stark localization \cite{wannier1962rmp}. Furthermore, it was recently proposed that for strong enough tilt the interacting model still features nonergodic behavior in the absence of disorder, an effect known as Stark MBL \cite{nieuwenburg2910pnas,schulz2019prl} and which is still the subject of debate \cite{kloss2022arxiv}.

\begin{figure*}[t]
\begin{center}
\includegraphics[scale=0.55]{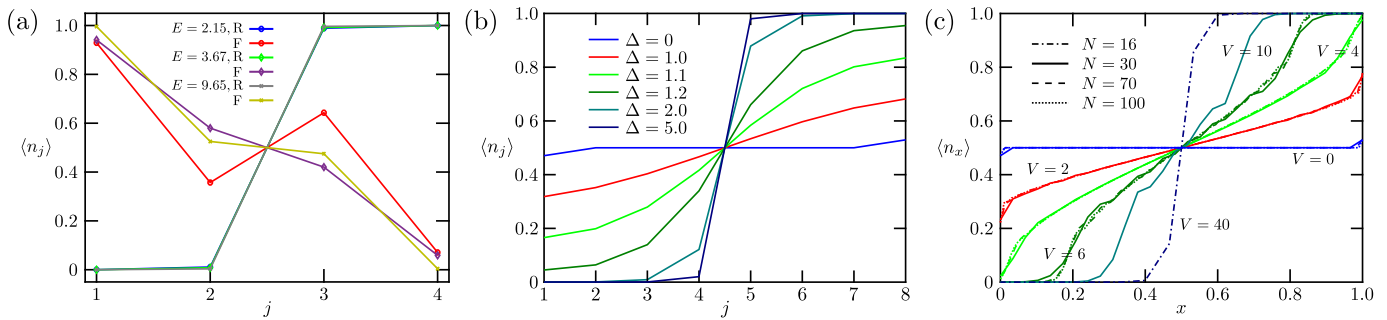}
\end{center}
\caption{Formation of particle domains. (a) Site population $\langle n_j\rangle$ for $N=4$, $\Delta=5$ and the resonant tilt values, for forward and reverse driving. (b) Site population $\langle n_j\rangle$ for a lattice of $N=8$, $E=0$ and different interactions $\Delta$ for reverse driving. (c) Site population $\langle n_x\rangle$ for $\Delta=0$, different system sizes and rescaled tilt $V$, with rescaled position $x = j/N \in[0,1]$.}
\label{fig_domains}
\end{figure*}

To drive the system into a nonequilibrium state, we incorporate high-temperature reservoirs with differing chemical potentials at its boundaries. We assume a weak coupling to the system (Born approximation), memory-less reservoirs (Markov approximation), and that the bandwidths of the reservoirs are much larger than those of the system, which leads to frequency-independent system-reservoir interactions (wide-band limit) \cite{mark2018njp}. Tracing out the reservoir degrees of freedom leads to a Lindblad master equation for the reduced density matrix $\rho$ of the system \cite{BreuerPetruccione},
\begin{align}\label{lindblad}
\frac{d\rho}{dt}=-i[H,\rho]+\sum_k\mathcal{L}_k(\rho),
\end{align}
with the dissipative superoperators $\mathcal{L}_k(\rho)$ defined by jump operators $L_k$ as
\begin{align}
\mathcal{L}_k(\rho)=L_k\rho L_k^{\dagger}-\frac{1}{2}\{L_k^{\dagger}L_k,\rho\}.
\end{align}
The driving is induced by boundary operators $L_{1,N}^+=\sqrt{\Gamma(1\pm f)/8}\ c_{1,N}^{\dagger}$ and $L_{1,N}^-=\sqrt{\Gamma(1\mp f)/8}\ c_{1,N}$, where $\Gamma$ is the coupling strength (taken as $\Gamma=1$) and the driving parameter $-1\leq f\leq1$ establishes the bias \cite{Benenti:2009,nosotros2013prb,landi2021arxiv}. Namely, when $f>0$, fermions are mostly created on site 1 and mostly annihilated on site $N$ giving forward driving, while inverting the sign of $f$ gives reverse driving. The particle current corresponds to the expectation value of the operator
\begin{align}
\mathcal{J}_j=i\frac{J}{2}\left(c_j^{\dagger}c_{j+1}-c_{j+1}^{\dagger}c_{j}\right),\qquad j=1,\ldots,N-1,
\end{align}
which directly arises from the particle number continuity equation \cite{landi2021arxiv}. We focus on the transport properties of the nonequilibrium steady state (NESS), where the current across the system is homogeneous, i.e. $\langle\mathcal{J}_j\rangle=\mathcal{J}$, as seen from the continuity equation. In addition, we set $|f|=1$ to consider maximal forward/reverse driving, and take $\mu=-E(N+1)/4$ so the system features charge conjugation and parity (CP) symmetry. This choice will help make the rectification mechanism transparent. Nonetheless, under the assumed wide-band limit for the driving, the transport is independent of $\mu$ \cite{znidaric2011pre} (see Appendix \ref{app_cpsymm}).

\subsection{Calculation of transport properties in interacting systems}
The transport properties of small lattices ($N\leq8$) were obtained exactly by diagonalizing the full Lindblad superoperator, written as a $4^{N}\times 4^{N}$ matrix. The NESS is the eigenstate with zero eigenvalue, reshaped as the $2^N\times 2^N$ density matrix $\rho$. For larger systems the calculation of the NESS is challenging. For forward transport, we describe $\rho$ as a matrix product state \cite{prosen2009jstat}, and obtain the particle current using the time evolving block decimation method \cite{VidalTEBD2004,VerstraeteTEBD2004} implemented with the Tensor Network Theory (TNT) library \cite{tnt,tnt_review1}. Here, the evolution of an arbitrary state was simulated until the current becomes homogeneous, indicating convergence. However, for strong tilts and interactions a very slow approach to the NESS was often encountered making full convergence across the system impractical. In such cases we applied the following strategy. From the application of the continuity equation to the boundaries of the system \cite{landi2021arxiv}, it is possible to relate the particle current to the population of the first site \cite{nosotros2013jstat}, so 
\begin{align} \label{curr_popul_general}
    \mathcal{J}=\frac{\Gamma}{8}\left(f+1-2\langle n_1\rangle\right).
\end{align}
Thus, converging the population of the first site leads to a direct calculation of the current. This shortcut allowed us to accurately reproduce the NESS current of small lattices justifying its application to longer chains.\\
For $N>8$ and reverse transport, where even the previous shortcut is insufficient due to exponentially slow convergence to the NESS \cite{Benenti:2009,nosotros2013prb}, we calculated the current from an analytical ansatz for insulating domain states. This is given by (see Appendix \ref{app_ansatz})
\begin{align} \label{ness_ansatz}
\rho_{\text{Dom}}=\sum_{n=0}^Np_n |\Psi^D_n\rangle\langle\Psi^D_n|,\qquad\sum_{n=0}^Np_n=1,
\end{align}
with perturbative domain eigenstates $|\Psi^D_n\rangle$ of each $n$ particle sector. The ansatz captures remarkably well the particle density profile allowing the current to be calculated from the population of the first site. Since the considered system sizes can have very small currents $\mathcal{J}<10^{-15}$, exact calculations from this ansatz were performed in {\tt MATLAB} using the variable-precision arithmetic (vpa) function to take 32 digits of precision.

\subsection{Rectification in small tilted systems}
The simple nonequilibrium scenario introduced at the start of Sec. \ref{sec_scheme} is so effective that large rectification can be observed even for very small lattices. For instance, in Fig. \ref{setup}(d) we illustrate the widely differing response of a system with $N=4$ and $\Delta=5$ when inverting the direction of the bias. For reverse driving of fermions down the tilted potential, the current decays rapidly and monotonically with $E$. On the other hand, for forward driving of fermions up the potential, the current features much larger values at well-defined resonances with $E$. As shown in Fig. \ref{setup}(e), this leads to rectification coefficients of up to $O(10^3)$, a value that can be vastly enhanced in longer chains (see Fig. \ref{fig_recti}).

\begin{figure*}[t]
\begin{center}
\includegraphics[scale=0.55]{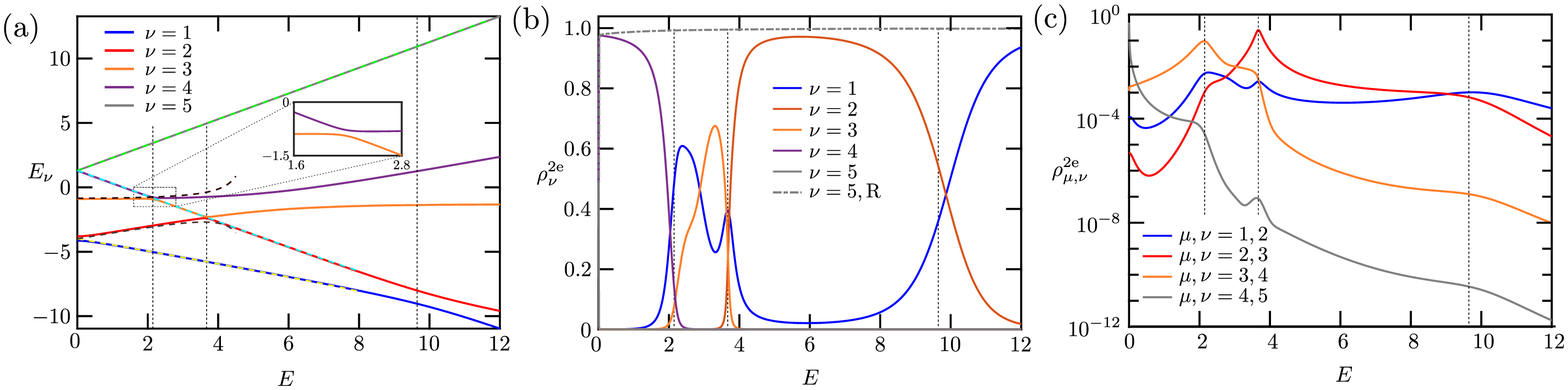}
\end{center}
\caption{Transport resonance mechanism in interacting tilted lattice of $N=4$ and $\Delta=5$, illustrated in the even CP symmetry sector of $n=2$ filling (sector $s=2\text{e}$). In all panels, the dotted vertical lines indicate the location of the forward current resonances shown in Fig. \ref{setup}(d). (a) Eigenenergies $E_{\nu}$ as a function of the tilt $E$. For low tilts, the eigenstate $\nu=4$ corresponds to the corrected left domain state. The inset amplifies the first avoided crossing. The dashed lines correspond to second order perturbative calculations of the eigenenergies (see Appendix \ref{app_pert_eigen}). (b) Probabilities $\rho^{2\text{e}}_{\nu}$ of eigenstates $\nu$ on the NESS. The line of $\nu=5,\text{R}$ corresponds to the highest energy eigenstate and reverse transport; all the others are for forward driving. Also note that the curve of $\nu=1$ is peaked at the first two resonances, even though it is not involved in an avoided crossing. This is because it corresponds to a perturbative correction of $|1010\rangle$ (see Appendix \ref{app_pert_eigen}), which is strongly coupled to the $f=1$ driving. (c) Coherences $\rho^{2\text{e}}_{\mu,\nu}$ of pairs of eigenstates $\mu,\nu$ for forward transport. At the avoided crossing of eigenstates $\mu$ and $\mu+1$, $\rho^{2\text{e}}_{\mu,\mu+1}$ is maximal, which breaks the insulating nature of the NESS. Also, when getting away from avoided crossings, the NESS is again mostly described by a single eigenstate, suppressing transport.}
\label{fig_reso_4sites}
\end{figure*}

These properties are intimately linked to the presence or absence of a particular distribution of fermions across the system, as depicted in Fig. \ref{fig_domains}(a). For reverse driving, the particles are pinned to the right boundary forming a domain half the size of the lattice, leaving the left half essentially unoccupied. The domain Pauli blocks more fermions from entering the system, resulting in an insulating state. On the other hand, for forward bias near the resonant values of $E$, the population profiles do not feature this constraint, so particles are allowed to move across the lattice. In the following Sections we explain in detail the emergence of these nonequilibrium effects.

\section{Reverse driving insulating NESS} \label{sec_insulating}
We consider first the insulating state. Domain formation is a common feature induced by strong interactions and tilt on their own, as depicted in Figs. \ref{fig_domains}(b) where $E=0$ and \ref{fig_domains}(c) where $\Delta = 0$. In both these scenarios there is no rectification; transport for forward and reverse drivings are identical up to a direction inversion, so we focus our attention to the latter. 

For strongly-interacting homogeneous ($E=0$) systems the insulating domain-like NESS is a well-known phenomenon \cite{Benenti:2009,nosotros2013prb,nosotros2013jstat,landi2021arxiv}, reproduced in Fig. \ref{fig_domains}(b) with exact diagonalization calculations for small lattices. The system features a flat profile in the noninteracting limit, typical of ballistic transport. As $\Delta$ increases so does the slope of the profile, until reaching extreme population values at the boundaries for $\Delta\gtrsim1$. Increasing $\Delta$ even further results in the creation of the particle domain. 

In contrast, noninteracting ($\Delta=0$) boundary-driven tilted systems have received much less attention \cite{landi2014pre,silva2022arxiv,de2022arxiv,pinho2023bloch} and their manifestation of expected localization has remained unexplored. We perform such an analysis obtaining exactly the site populations $\langle n_j\rangle$ and particle current $\mathcal{J}$ for large systems by reducing the Lindblad master equation \eqref{lindblad} to a set of linear equations (see Appendix \ref{app_noninter}). We use a rescaled tilt $V=EN$, to keep the on-site energy difference between boundaries approximately constant. This way the density profiles of different system sizes collapse for each $V$, evidencing the emergence of domains for small lattices and large tilt, as well as for long lattices and low tilt (i.e. when the total tilt is larger than the Bloch bandwidth, $V\gtrsim4$ \cite{de2022arxiv}). In addition, for large systems the particle current decays exponential with $N$ and $V$, as shown in Fig. \ref{delta0_transport} of Appendix \ref{app_noninter}.

While the physical origin of the interaction- and tilt-induced insulating states is different, they share the same underlying mechanism that remains applicable for tilted interacting lattices. Specifically, they arise from the interplay between the gapped eigenstructure of the system and the maximal boundary driving. For small hopping $J\ll \Delta\text{ or }E$, the highest energy eigenstate $|\Psi^D_{n}\rangle$ of each $n$ particle sector (e.g. eigenstate $\nu=5$ of Fig. \ref{fig_reso_4sites}(a)) corresponds to a perturbative correction to the configuration state $|00\ldots0_{N-n}11\ldots1_n\rangle$ in which all $n$ particles form a domain pinned to the right. The energy gap between this eigenstate and the rest, which increases with $\Delta$ and $E$, ensures that the amplitudes of configuration states in $|\Psi^D_{n}\rangle$ that couple to the $f=-1$ driving (where a particle can be injected at site $N$ or ejected at site $1$) are exponentially suppressed (see Appendix \ref{app_ansatz}). Thus the eigenstates $|\Psi^D_{n}\rangle$ become exponentially close to obeying
\begin{align}
\sum_k\mathcal{L}_k(|\Psi^D_{n}\rangle\langle\Psi^D_{n}|)= 0,
\end{align}
making them {\em dark states} of the driving. The overall NESS density matrix $\rho$ of the system is very well captured by an analytical ansatz built from a statistical mixture of these darks states introduced already in Eq. \eqref{ness_ansatz}. From this, we show that $\rho$ is strongly dominated by the contribution $|\Psi^D_n\rangle\langle\Psi^D_n|$ with $n=N/2$, corresponding to a state with occupied and empty domains of equal size (e.g. near unit probability for eigenstate $\nu=5$ for reverse driving in Fig. \ref{fig_reso_4sites}(b)). These results directly lead to the insulating NESSs of equal empty and occupied domains for systems with strong interactions \cite{nosotros2013prb}, tilt or both, whose currents monotonically and exponentially decrease as either increase.\\

\section{Forward driving resonances} \label{sec_resonances}
Next we focus on the forward driving setup with $f=1$. Here, strong interactions and tilt on their own would favor a NESS largely dominated by perturbative corrections to the left domain $|11\ldots1_{\frac{N}{2}}00\ldots0_{\frac{N}{2}}\rangle$. However, this configuration is the highest energy state of the $n=N/2$ particle number sector (degenerate with the right domain) for an interaction-dominated scenario, while it is the lowest energy state for a strongly tilted lattice. Thus, increasing $\Delta$ raises the energy of this state, while increasing $E$ lowers it. 
For instance, fixing $\Delta$ and sweeping through $E$ (as in Fig. \ref{fig_reso_4sites}(a)) closes the energy gap from the initial NESS dark state, eventually forcing it to cross through the bulk spectrum of the system (see sketch in Fig. \ref{setup}(c)). As a result, current resonances are induced. This effect is cleanly illustrated by the half-filled even CP symmetric sector with the same parameters of Fig. \ref{setup}(d) as shown in Fig. \ref{fig_reso_4sites}(a). Crucially, since all the eigenstates have the same symmetry, the no-crossing theorem \cite{Landau} ensures that only avoided crossings occur. Their location accurately pinpoints the current resonances. 

\begin{figure*}[t]
\begin{center}
\includegraphics[scale=0.55]{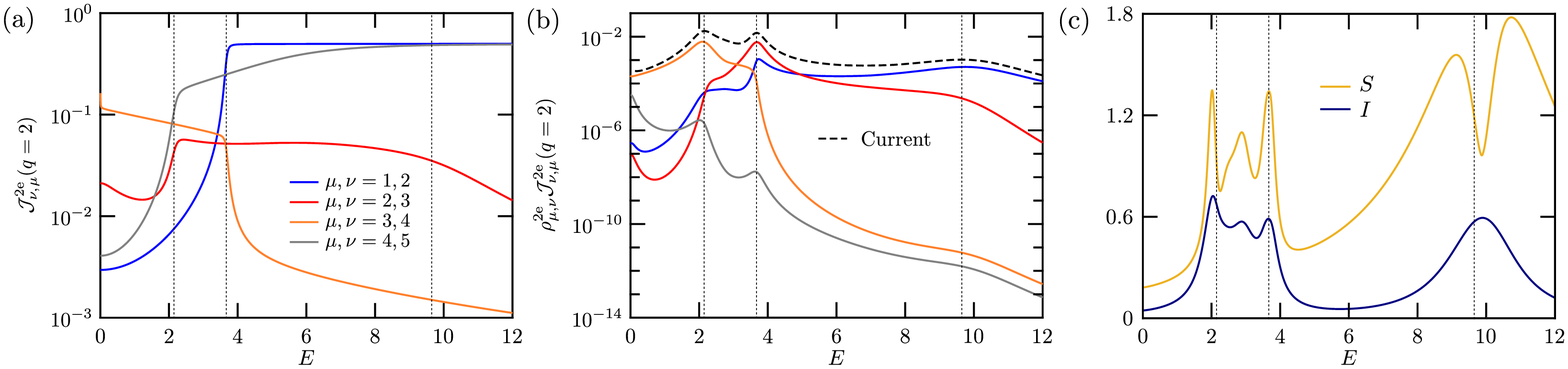}
\end{center}
\caption{(a) Current matrix elements $\mathcal{J}^{2e}_{\nu,\mu}(q=2)$ for interacting tilted system of $N=4$ and $\Delta=5$. (b) Product of coherences $\rho^{2e}_{\mu,\nu}$ and matrix elements $\mathcal{J}^{2e}_{\nu,\mu}(q=2)$, which gives a contribution from the half-filled even CP symmetry sector to the total current (dashed line). (c) Impurity ($I$) and OSEE ($S$). The second peak in both quantities, located between the first and second current resonances, emerges from the high population of eigenstates $\nu=1,3$ (see Fig. \ref{fig_reso_4sites}(b)). In all panels, the dotted vertical lines indicate the location of the forward current resonances, corresponding to the peaks of the forward current depicted in Fig. \ref{setup}(d)}.
\label{fig6sm}
\end{figure*}

To understand the role of the avoided crossings we evaluate the contributions to the particle current in the energy eigenbasis. Since there are no coherences between different symmetry sectors, the current is given by 
\begin{align}
\mathcal{J}=\sum_{s}\sum_{\mu_s,\nu_s}\rho^s_{\mu_s,\nu_s}{J}^{s}_{\nu_s,\mu_s}(q),
\end{align}
where $s$ sums over all the sectors, $\rho^s_{\mu_s,\nu_s}=\langle\mu_s|\rho|\nu_s\rangle$ is the coherence between eigenstates $|\mu_s\rangle$ and $|\nu_s\rangle$ of sector $s$, and ${J}^{s}_{\nu_s,\mu_s}(q)=\langle\nu_s|\mathcal{J}_q|\mu_s\rangle$ is the corresponding current matrix element evaluated on some site $q$. From the probabilities $\rho^s_{\nu_s}=\rho^s_{\nu_s,\nu_s}$ and coherences $\rho^s_{\mu_s,\nu_s}$, shown in Figs. \ref{fig_reso_4sites}(b) and \ref{fig_reso_4sites}(c) respectively, the following picture emerges. For very low tilts and strong interactions, the NESS is solely dominated by the high-energy left domain-like state, evidenced by its large probability. This eigenstate has an energy $\sim\frac{1}{4}\left[\Delta(N-3)?En(N-n)\right]+\mathcal{O}(J^2)$ to lowest perturbative order (see Appendix \ref{app_pert_eigen}), so as $E$ increases it dives down towards the lower eigenenergies which have a weaker dependence on the tilt. As it approaches another eigenstate they mix and probability is transferred from the higher to the lower energy states. Thus, in the proximity of the avoided crossing the Hamiltonian no longer supports an eigenstate that is dark to the $f=1$ driving. This allows large coherences to develop, enhancing the current. Further avoided crossings occur for even larger $E$, with lower eigenstates also getting populated. Eventually for very large $E$ the whole spectrum is crossed and the lowest-energy eigenstate $\nu=1$ inherits the dominant role on the NESS. This leads again to an insulating left domain-like dark state, but now induced by the strong tilt.

Notably, this picture is further evidenced by the behavior of the population profiles of the Hamiltonian eigenstates as the tilt is varied, discussed in detail in Appendix \ref{app_eigen_popul}. Overall, these results show that away from the resonances the eigenstate of highest probability at the NESS corresponds to a left domain, while close to the resonances the domains are suppressed and the eigenvectors involved in the corresponding avoided crossing exchange the form of their population profile.

Some important remarks are in order regarding the transport resonance mechanism underlying our quantum diode. First, such mechanism is not restricted to half filling; different particle number sectors feature avoided crossings corresponding to forward current maxima, which might even be absent from the half-filled sector. Considering this, we have accounted for every resonance reported in this work. Second, in addition to being associated to a large coherence, a forward current resonance takes place at an avoided crossing if the corresponding current matrix element ${J}^{s}_{\nu_s,\mu_s}(q)$ is significant. Otherwise, the contribution of that coherence to the current is small. This is already evidenced from the coherences shown in Fig. \ref{fig_reso_4sites}(c), and the current matrix elements of Fig. \ref{fig6sm}(a). The coherence between states $\mu,\nu=3,4$ has a large current matrix element at the tilt of the first resonance; the same holds for the coherence between states $\mu,\nu=2,3$ at the second resonance, and for that between states $\mu=1,2$ at the third resonance. Because of this, the product $\rho^{2e}_{\mu,\nu}\mathcal{J}^{2e}_{\nu,\mu}(q=2)$ is large, as depicted in Fig. \ref{fig6sm}(b), and a resonance is clearly seen. On the other hand, even though there is a large coherence between states $\mu,\nu=4,5$ for $E\ll1$, the corresponding current matrix element is strongly suppressed, leading to a barely-seen current maximum around $E\approx0$. In conclusion, not every large coherence, arising e.g. from an avoided crossing, will correspond to a forward transport resonance. 

Finally, to emphasize the many-body complexity of the NESS at forward transport, we calculate the impurity $I$ of the full NESS,
\begin{align}
I=1-\text{tr}\left(\rho^2\right),
\end{align}
and the operator space entanglement entropy (OSEE), \cite{prosen2007pra,prosen2009jstat,nosotros2013prb},
\begin{align}
S=-\sum_{\alpha}\lambda_{\alpha}^2\log_2\lambda_{\alpha}^2,
\end{align}
where $\lambda_{\alpha}$ are the Schmidt coefficients of the half-size bipartition of the system. The former indicates how much the NESS deviates from a pure state. The latter has been well established as a measure of the complexity of quantum operators \cite{alba2019prl,bertini2020scipost,wellnitz2022prl,gong2022prl} (here a density matrix $\rho$ of a dissipative many-body quantum system). Furthermore, being directly obtained from the singular value decomposition of the matrix product state representation of $\rho$, it quantifies the computational effort in capturing the nonequilibrium dynamics of the open system with tensor networks methods.

\begin{figure*}[t]
\begin{center}
\includegraphics[scale=0.52]{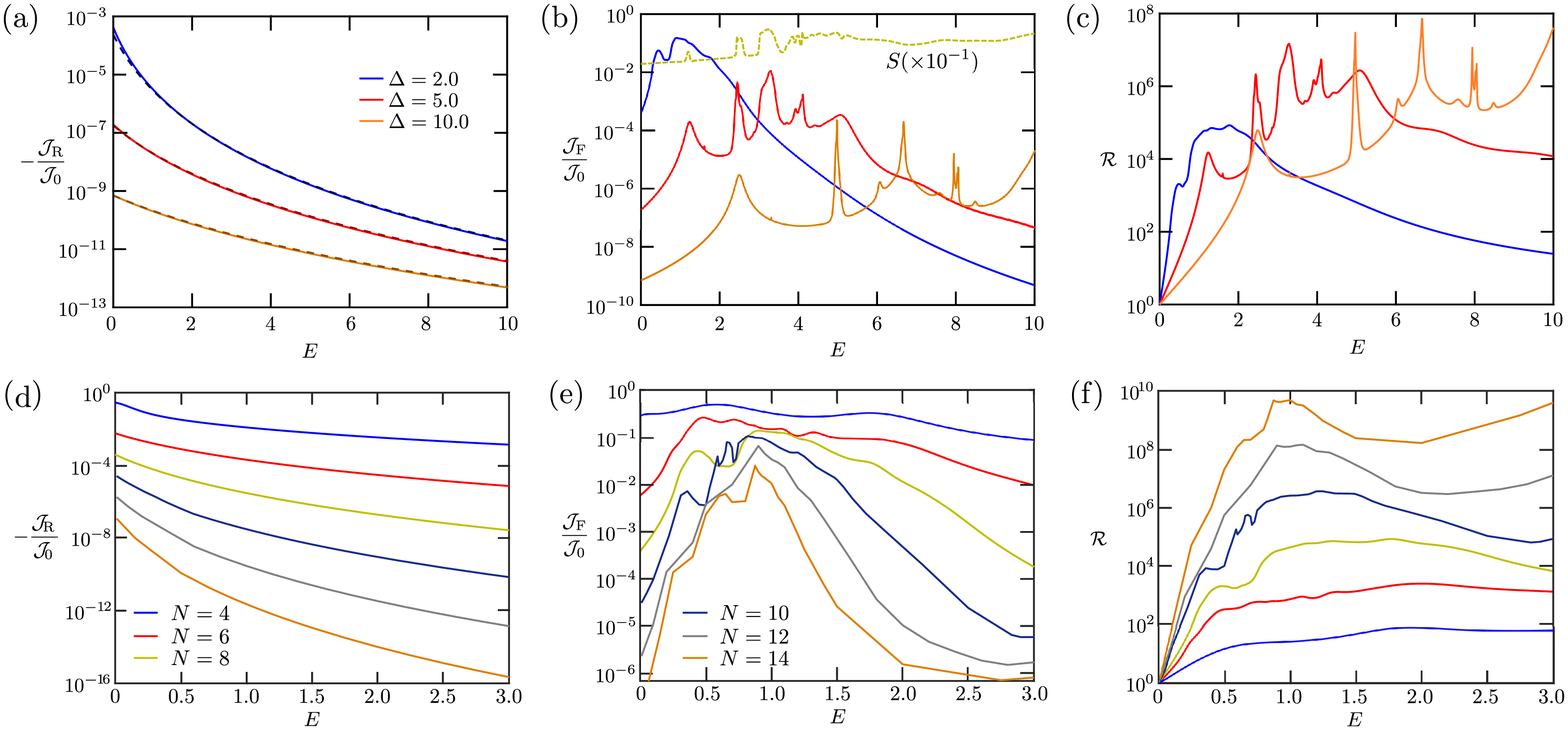}
\end{center}
\caption{Transport and rectification in larger interacting tilted systems. (a) Reverse particle current $\mathcal{J}_\text{R}$ as a function of tilt for $N=8$ and different interaction strengths. The solid lines correspond to exact diagonalization results, and the darker dashed lines to currents obtained from our analytical Ansatz (see Appendix \ref{app_ansatz}). (b) Exact forward particle current $\mathcal{J}_\text{F}$ for the same parameters. The dashed line represents the OSEE (divided by $10$) for the system with $\Delta=5$. (c) Corresponding rectification coefficient $\mathcal{R}$. (d) Reverse particle current $\mathcal{J}_\text{R}$ as a function of tilt for $\Delta=2$ and different system sizes. Given that the analytical Anstatz slightly deviates from the exact results at low tilts $E\leq0.5$, as seen in panel (a) for $N=8$, we obtain the reverse currents of $N>8$ in that regime from matrix product state simulations. (e) Forward particle current $\mathcal{J}_\text{F}$ for the same parameters. (f) Corresponding rectification coefficient $\mathcal{R}$. The legends of system sizes spread over (d) and (e) apply to the three bottom panels. Also, the currents have been rescaled by the maximal current $\mathcal{J}_0$ obtained with the same driving for a noninteracting homogeneous system (see Eq. \eqref{curr_noE}).}
\label{fig_recti}
\end{figure*}

Both quantities are plotted in Fig. \ref{fig6sm}(c), and feature maximal values at or very close to the current resonances. Intuitively we expect this to be the case: far from the resonances, where only one energy eigenstate is largely populated (so $\rho$ is close to being pure), such a state approximately corresponds to a left product domain (low OSEE). Close to the resonances, the mixing of energy eigenstates naturally reduces the purity and also leads to the emergence of strong correlations, increasing the complexity of a matrix product state representation of $\rho$. These results reinforce that the observed enhanced conduction is a collective nonequilibrium phenomenon in an interacting many-body system.

\section{Giant rectification} \label{sec_recti}

The combination of the enhanced insulating state of reverse driving in Sec. \ref{sec_insulating} with the forward resonant behavior of Sec. \ref{sec_resonances} results in sizable rectification coefficients $\mathcal{R}=-\mathcal{J}_\text{F}/\mathcal{J}_\text{R}$ even for the small chain of the example (see Fig. \ref{setup}(e)). Crucially, rectification becomes more substantial and more robustly achievable for larger system sizes $N$. The enhancement in the maximum $\mathcal{R}$ with $N$ follows from the differing current scaling of reverse and forward drivings. Namely $\mathcal{J}_{\rm R} \sim \exp(-N)$, characteristic of an insulator, while at a resonance we expect $\mathcal{J}_{\rm F} \sim N^{-\alpha}$ with some exponent $\alpha>0$ characteristic of a conductor \cite{bertini2021rmp,landi2021arxiv}. As shown in Fig. \ref{fig_recti}(c), already for $N=8$ with $\Delta=5$ we find coefficients of up to $O(10^7)$ compared to $O(10^3)$ for $N=4$. In addition, $\mathcal{R}$ is further enhanced by particle interactions. As depicted in Figs. \ref{fig_recti}(a)-(c), while increasing $\Delta$ diminishes both reverse and forward currents, the former decrease is much more prominent and thus their ratio features even higher resonances. Notably, here the current resonances also agree with maximal values of measures of complexity of many-body dynamics, as exemplified in Fig. \ref{fig_recti}(b) with the OSEE as a function of $E$ for $\Delta=5$. 

While in small systems achieving the maximum $\mathcal{R}$ requires fine-tuning of $E$ to locate a resonance, for larger $N$ this constraint is rapidly removed due to the exponentially increasing number of eigenstates crossed within the bulk of the spectrum. As a result the rectification is broadened over the many-body bandwidth of the chain. Moreover, the favourable current scaling with $N$ means that giant rectification can be achieved with a reduced interaction strength. To evidence these effects, using the insulating ansatz and tensor network simulations, we calculated reverse and forward currents for $\Delta=2$ and systems of $N>8$, shown in Figs. \ref{fig_recti}(d) and \ref{fig_recti}(e) respectively. From these we find $\mathcal{R}\sim\mathcal{O}(10^{9})$ over a range of $E$, as reported in Fig. \ref{fig_recti}(f). This establishes larger rectification coefficients than those previously obtained for different types of non-homogeneous lattices \cite{balachandran2018prl,lee2020entropy,lee2021pre,poulsen2022pra,lee2022pre}, with a more intuitive and easily implementable setup \cite{guardado2020prx,scherg2021nat,hebbe2021prx,guo2021prl}. 

Finally, we note that the giant rectification effect discussed here is not an artefact of the driving scheme; it also emerges in more realistic boundary-driven setups. The Lindblad driving directly applied to the boundary sites of the system corresponds to fermionic baths with a flat and extremely wide spectral function at high-temperature \cite{Benenti:2009}. To move away from this idealized limit we explore in Appendix \ref{app_mesoreservoirs} a \textit{mesoscopic leads} approach. Here the infinite reservoirs at given temperatures and chemical potentials are replaced by leads consisting of a finite set of damped fermionic modes \cite{nosotros2020prx,marlon2023prb,de2022arxiv}. We consider a simple but non-trivial driving scheme with leads comprising two modes that represent a double Lorentzian bath spectral function at high temperature and strong bias. As depicted in Fig. \ref{fig_mesoreservoirs}, even larger peaks of the rectification coefficient compared to those of Fig. \ref{setup} emerge, with both taking place at the same tilt values. Thus, we conclude that our giant-rectification quantum diode mechanism is implementable in more realistic nonequilibrium scenarios.

\section{Conclusions and Outlook} \label{sec_conclu}
We have proposed a giant rectification scheme in correlated quantum systems, based on the natural asymmetry between forward and reverse transport in tilted interacting lattices. In our protocol, particles get locked into an insulating domain NESS when attempting to travel down the tilted potential (reverse bias). On the other hand, they propagate uphill resonantly as avoided crossings in the spectrum are approached (forward bias). Our calculations show rectification coefficients of up to $O(10^{9})$ for small chains, a number that can be pushed even further without fine tuning and at moderate interaction strengths by considering larger systems.

We emphasize that most of the reported rectification peaks are associated to sizeable values of the forward particle currents. This is demonstrated in Figs. \ref{setup} and \ref{fig_recti} by rescaling $\mathcal{J}$ with the maximal current $\mathcal{J}_0$ achievable under the same driving scheme for a noninteracting lattice with no tilt, which is a well-known size-independent ballistic current. Consequently at the resonances the system has sizeable conduction in the forward direction while it is insulating in reverse bias. Thus the observed giant rectification is not an artifact of dividing two very small currents associated with insulating states \cite{haack2023prr}.

Our simple yet rich model serves as an ideal framework to unravel further nontrivial many-particle effects out of equilibrium. Namely, our results pave the way for developing different types of giant rectification protocols (e.g. of spin, charge or heat) in a variety of driving schemes, including finite temperature reservoirs, magnetic or thermal biases, and even excitation by light \cite{boolakee2022nat}. For this purpose, recent approaches for accurately simulating more realistic reservoirs, even when they are strongly coupled to the system, might be exploited \cite{rams2020prl,nosotros2020prx,archak2021prb,fux2021prl,riera2021prx,marlon2023prb,de2022arxiv}. In addition, our work motivates further analysis on the nature of transport in the forward regime. This includes numerical simulations of larger driven systems, as well as analytical calculations of operators that overlap with the current based on the concept of dynamical $l-$bits \cite{berislav2021arxiv,berislav2022arxiv}. Furthermore, our proposal could be realized with existing electronic setups and quantum simulators. On the one hand, a tilted potential can be directly created on an electronic device coupled to metallic leads by an electric field \cite{boolakee2022nat}. On the other, several experiments have implemented trapped-ion \cite{morong2021nat}, cold-atom \cite{guardado2020prx,scherg2021nat,hebbe2021prx} and superconducting qubit \cite{guo2021prl} setups with a similar tilt and/or interactions to those used in the present work, and even in larger systems. Current quantum computing devices also constitute platforms where our scheme can be implemented \cite{google2023arxiv,erbanni2023arxiv}. In addition, exploiting boundary-driven architectures in cold-atom platforms \cite{pepino2009prl,krinner2017jpcm,amico2022arxiv} or reservoir engineering for irreversible particle injection in superconducting circuits \cite{ma2019nat}, similar versions of our diode could be incorporated in state-of-the-art nanoscale quantum circuits.  

\section*{Acknowledgments}
The authors thank B. Bu\v{c}a for his comments on the manuscript, and D. Jaksch for insightful discussions. The authors also gratefully acknowledge financial support from UK's Engineering and Physical Sciences Research Council (EPSRC) under grant EP/T028424/1. This work is part of an effort sponsored by AFOSR Grant FA9550-23-1-0014.

\appendix

\section{Transport in noninteracting boundary-driven tilted systems} \label{app_noninter}
In this Appendix we describe our approach to obtain the transport properties of non-interacting boundary-driven tilted systems. Even though we do not obtain the full solution for the density matrix, the relevant observables are calculated exactly up to numerical precision. For this, we propose the ansatz for the density matrix \cite{znidaric2010jstat,znidaric2011pre}
\begin{align} \label{ansatz}
\begin{split}
\rho&=\frac{1}{2^N}I+\sum_{k=1}^N2a_k\left(n_k-\frac{1}{2}\right)\\
&+\sum_{l=2}^N\sum_{k=1}^{N-l+1}b_k^{(l)}B_k^{(l)}+\sum_{l=2}^N\sum_{k=1}^{N-l+1}h_k^{(l)}H_k^{(l)}+\cdots
\end{split}
\end{align}
where we define the $l-$site operators
\begin{align}
\begin{split}
&B_k^{(l)}=-2i\left(c_j^{\dagger}c_{j+l-1}-c_{j+l-1}^{\dagger}c_{j}\right)\\
&H_k^{(l)}=2\left(c_j^{\dagger}c_{j+l-1}+c_{j+l-1}^{\dagger}c_{j}\right),
\end{split}
\end{align}
which correspond to generalized current and energy density operators respectively; in particular, $B_k^{(2)}=-4\mathcal{J}_k$. Even though this is a first-order expansion in $f$ \cite{znidaric2010jstat,znidaric2011pre}, it leads to an exact solution of the parameters $a_k=\langle2n_k-1\rangle$, $b_k^{(l)}=\langle B_k^{(l)}\rangle/2$ and $h_k^{(l)}=\langle H_k^{(l)}\rangle/2$. For this, the ansatz \eqref{ansatz} in inserted into the master equation \eqref{lindblad} with the stationary condition $d\rho/dt=0$, which results in a closed set of linear equations for the parameters when grouping the coefficients in front of each operator.

\subsection{Steady state equations}
Importantly, due to the symmetric boundary driving several coefficients of the ansatz are zero or equal to others. Assuming $N$ even, the symmetry relations are
\begin{align}
a_k=-a_{N-k+1},\qquad &k=1,\ldots,\frac{N}{2},    
\end{align}
for the populations and 
\begin{align}
h_k^{(l)}=(-1)^lh_{N-l-k+2}^{(l)},\qquad b_k^{(l)}=(-1)^lb_{N-l-k+2}^{(l)},&
\end{align}
with $k=1,\ldots\frac{N-2l'}{2}+1$, and where $l=2l'$ for $l$ even and $l=2l'-1$ for $l$ odd. If $N$ is odd the central population vanishes, namely $a_{\frac{N+1}{2}}=0$, the recursions for $h_k^{(l)}$ and $b_k^{(l)}$ hold but for $k=1,\ldots\frac{N-2l'+1}{2}$, and 
\begin{align}
h_{\frac{N-l}{2}+1}^{(l)}=b_{\frac{N-l}{2}+1}^{(l)}=0\text{ for }l\text{ odd.}
\end{align}
Thus with $N$ even we have $\frac{N^2}{2}+1$ unknown coefficients, while for $N$ odd we have $\frac{(N-1)^2}{2}+1$ unknown coefficients and $N$ zero coefficients. This indicates that the same effort is required to solve systems of $N=2M$ and $N=2M+1$ sites.\\
Considering these symmetries, we obtain the following set of equations after inserting the ansatz in the master equation and grouping the coefficients of each resulting operator. From the coefficients in front of $2n_1-1$ we get
\begin{align} \label{relation_a1_b}
a_1+\frac{4b^{(2)}}{\Gamma}=f,\qquad\text{with }b^{(2)}=b_k^{(2)}\text{ homogeneous}.
\end{align}
This equation, consistent with Eq. \eqref{curr_popul_general}, establishes a direct link between the boundary population and the particle current, allowing us to obtain one from the other. \\
\begin{figure*}[t]
\begin{center}
\includegraphics[scale=0.54]{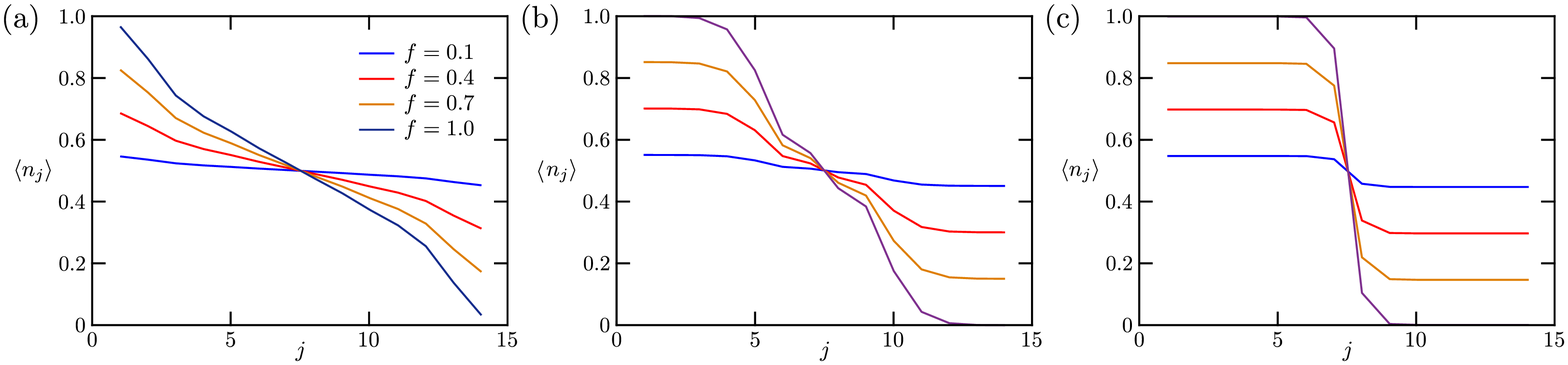}
\end{center}
\caption{Population profiles for noninteracting lattices with $N=14$, different tilts $E$ and driving parameters $f$. (a) $E=0.3$. (b) $E=0.6$. (c) $E=3.0$.}
\label{delta0_populations}
\end{figure*}
From the coefficients in front of $B_k^{(2)}$ ($k=1,\ldots,\frac{N}{2}-1$) we have
\begin{align} \label{eqBk2}
\begin{split}
-\frac{\Gamma}{2}b^{(2)}\delta_{k,1}&+2J(a_k-a_{k+1})\\&+2J\left(h_{k-1}^{(3)}-h_k^{(3)}\right)-2Eh_k^{(2)}=0.
\end{split}
\end{align}
This equation is modified for $k=\frac{N}{2}$ (i.e. when reaching the center of the system), since $a_{\frac{N}{2}}=-a_{\frac{N}{2}+1}$ and $h_{\frac{N}{2}}^{(3)}=-h_{\frac{N}{2}-1}^{(3)}$. Thus here we have
\begin{align} \label{change_jk}
2Ja_{\frac{N}{2}}+2Jh_{\frac{N}{2}-1}^{(3)}-Eh_{\frac{N}{2}}^{(2)}=0.
\end{align}
From the coefficients in front of $H_k^{(2)}$ ($k=1,\ldots,\frac{N}{2}-1$) we have
\begin{align}
-\frac{\Gamma}{2}h_1^{(2)}\delta_{k,1}+2J\left(b_k^{(3)}-b_{k-1}^{(3)}\right)+2Eb^{(2)}=0,
\end{align}
which changes for $k=\frac{N}{2}$ since $b_{\frac{N}{2}}^{(3)}=-b_{\frac{N}{2}-1}^{(3)}$, giving
\begin{align} \label{change_h2k}
-4Jb_{\frac{N}{2}-1}^{(3)}+2Eb^{(2)}=0.
\end{align}
For $l\geq3$ even, from the coefficients in front of $B_k^{(l)}$ ($k=1,\ldots,\frac{N-l}{2}$), we get
\begin{align}\label{reduced_bl}
\begin{split}
&-\frac{\Gamma}{2}b_1^{(l)}\delta_{k,1}\left(1+\delta_{lN}\right)+2J\left(h_{k+1}^{(l-1)}-h_{k}^{(l-1)}\right)\\&+2J\left(h_{k-1}^{(l+1)}-h_k^{(l+1)}\right)-2(l-1)Eh_k^{(l)}=0,
\end{split}
\end{align}
and from the coefficients in front of $H_k^{(l)}$ we have
\begin{align}\label{reduced_hl}
\begin{split}
&-\frac{\Gamma}{2}h_1^{(l)}\delta_{k,1}\left(1+\delta_{lN}\right)+2J\left(b_{k}^{(l-1)}-b_{k+1}^{(l-1)}\right)\\&+2J\left(b_{k}^{(l+1)}-b_{k-1}^{(l+1)}\right)+2(l-1)Eb_k^{(l)}=0,
\end{split}
\end{align}
where the $\delta_{lN}$ results because for the last $l$ the driving terms of both boundaries are equal and thus are summed. For $k=\frac{N-l}{2}+1$, the corresponding equations are
\begin{align}
-&2Jh_{\frac{N-l}{2}+1}^{(l-1)}+2Jh_{\frac{N-l}{2}}^{(l+1)}-(l-1)Eh_{\frac{N-l}{2}+1}^{(l)}=0,\\
&2Jb_{\frac{N-l}{2}+1}^{(l-1)}-2Jb_{\frac{N-l}{2}}^{(l+1)}+(l-1)Eb_{\frac{N-l}{2}+1}^{(l)}=0.
\end{align}
Finally, for $l\geq3$ odd, the same relations hold for $k=1,\ldots,\frac{N-l+1}{2}$. This completes the set of $\frac{N^2}{2}+1$ linear equations for $N$ even and of $\frac{(N-1)^2}{2}+1$ equations for $N$ odd, which can be solved exactly for hundreds of sites with standard linear algebra packages and a moderate computational effort.\\
Importantly, if $E=0$, this problem reduces to that of a noninteracting ballistic conductor \cite{znidaric2010jstat,znidaric2011pre}. Here the equations for coefficients in front of $H_k^{(2)}$, $H_k^{(l)}$ and $B_k^{(l)}$ for $l\geq3$ decouple from the rest, and form a set of homogeneous equations with the same number on unknowns; the solution of this system is the trivial one where the corresponding coefficients $h_k^{(2)}$, $h_k^{(l)}$ and $b_k^{(l)}$ are zero. This leads to the size-independent particle current
\begin{align} \label{curr_noE}
\mathcal{J}_0=\frac{2\Gamma f}{\Gamma^2+16},
\end{align}
and a flat population profile in the bulk, as seen from Eq. \eqref{eqBk2}. In addition, if $E\neq0$ but $f=0$, the full system of equations is homogeneous, where only the trivial solution of all parameters being zero is possible. This corresponds to the NESS being equal to the identity (infinite temperature state), as expected.

\subsection{Analytical results for small chains}
Now we provide analytical results for the exact set of linear equations of $\Delta=0$, focusing on the particle current and population profile for small lattices. We take first $N=4$, which corresponds to a set of $9$ equations and $9$ unknowns, namely $a_1,a_2,b^{(2)},h_1^{(2)},h_2^{(2)},b_1^{(3)},h_1^{(3)},b_1^{(4)}$ and $h_1^{(4)}$. Even though this system can be easily solved, it leads to cumbersome expressions for the transport properties. Thus we consider two limits. First, for $E\ll1$ the particle current to lowest order in $E$ is given by
\begin{align} \label{curr_smallE_N4}
\mathcal{J}=\frac{2\Gamma f}{\Gamma^2+16}-\frac{1}{2}E^2\Gamma f\frac{(160+\Gamma^2)}{(16+\Gamma^2)^2}.
\end{align}
Thus the current decreases quadratically from the $E=0$ result of Eq. \eqref{curr_noE}. A more interesting situation is that of very large tilt $E\gg1$. Here we get the current
\begin{align} \label{curr_large_E}
\mathcal{J}=\frac{1}{8}\Gamma f E^{-4}
\end{align}
and the coefficients
\begin{align} \label{a_values_N4}
a_1=f(1-E^{-4}),\qquad a_2=f\left(1-\frac{5}{2}E^{-2}\right),
\end{align}
with $a_3=-a_2$ and $a_4=-a_1$. This indicates that a large tilt leads to an insulating state. The current is suppressed as a power law of $E$, as well as the deviations of the populations from $(1+f)/2$ and $(1-f)/2$ on the left and right halves of the chain, respectively. Indeed, if $f=1$, as in our numerical calculations, the populations $\langle n_k\rangle=(a_k+1)/2$ of the chain are
\begin{align}
\begin{split} \label{popul_N4_analytic_D0}
    &\langle n_1\rangle=1-\frac{1}{2}E^{-4},\qquad\langle n_2\rangle=1-\frac{5}{4}E^{-2},\\
    &\langle n_3\rangle=\frac{5}{4}E^{-2},\qquad\langle n_4\rangle=\frac{1}{2}E^{-4}.
\end{split}
\end{align}
This corresponds to the formation of almost fully-occupied and empty domains on the left and right halves of the system respectively. If $f<1$ the system is also insulating, with domains of population values closer to $1/2$, as shown in Fig. \ref{delta0_populations} for larger systems.

In addition, we calculated the exact solution of the linear set of equations for $N=5$, which has the same number of unknowns. In the limit of low tilt $E\ll1$, we get
\begin{align} \label{curr_smallE_N5}
\mathcal{J}^{\text{S}}=\frac{2\Gamma f}{\Gamma^2+16}-2E^2\Gamma f\frac{(80+\Gamma^2)}{(16+\Gamma^2)^2}.
\end{align}
It has a very similar form to that of $N=4$, but with larger deviation from the ballistic result. On the other hand, in the limit $E\gg1$, we get the current
\begin{align} \label{currN5_large_E}
\mathcal{J}=\frac{1}{32}\Gamma f E^{-4}
\end{align}
and $f=1$ populations
\begin{align}
\begin{split} \label{popul_N5_analytic_D0}
    &\langle n_1\rangle=1-\frac{1}{8}E^{-4},\qquad\langle n_2\rangle=1-\frac{1}{2}E^{-2},\\
    &\langle n_3\rangle=0,\qquad\langle n_4\rangle=\frac{1}{2}E^{-2},\qquad\langle n_5\rangle=\frac{1}{8}E^{-4}.
\end{split}
\end{align}
\begin{figure*}[t]
\begin{center}
\includegraphics[scale=0.75]{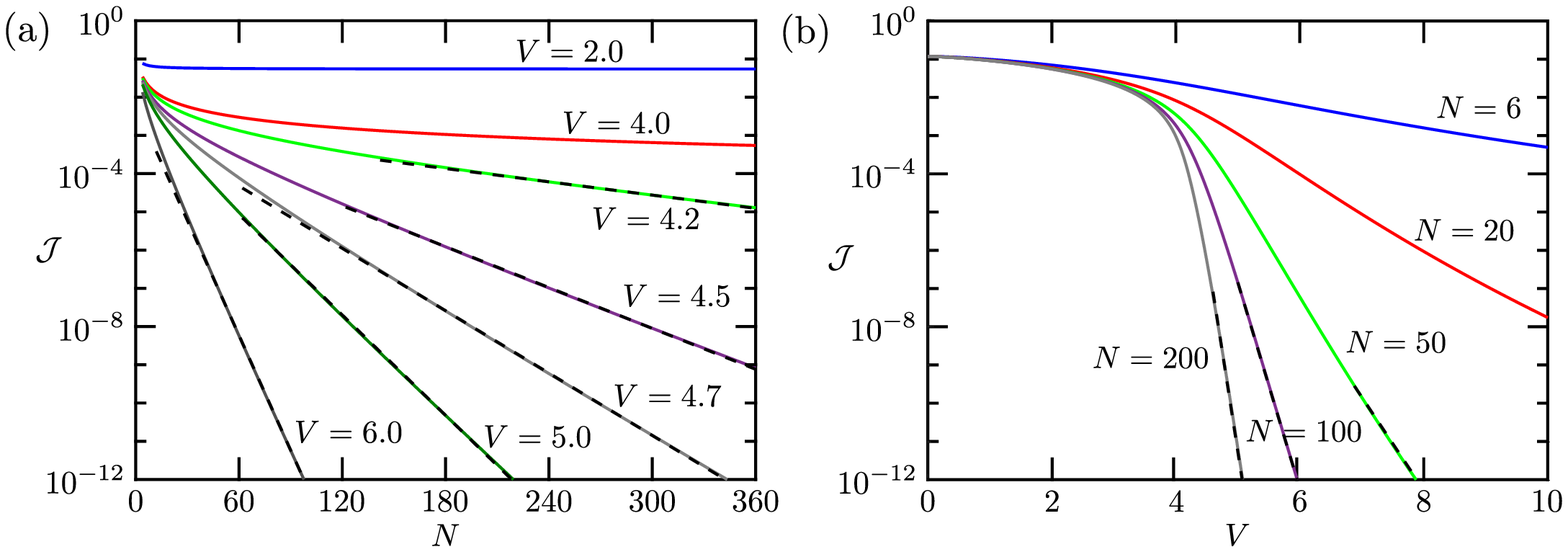}
\end{center}
\caption{Particle current in noninteracting tilted lattices. (a) As a function of $N$ for different values of $V$. The dashed lines correspond to fits to exponential functions $\sim e^{-cN}$, with exponents $c=0.0133(1),0.0410(2),0.0626(2),0.101(1),0.231(1)$ for $V=4.2,4.5,4.7,5.0,6.0$ respectively. (b) As a function of $V$ for different values of $N$. The dashed lines correspond to fits to exponential functions $\sim e^{-dV}$, with exponents $d=5.60(5),12.42(6),24.1(3)$ for $N=50,100,200$ respectively.}
\label{delta0_transport}
\end{figure*}
Thus in the large $E$ limit the current for $N=5$ is four times lower than that of $N=4$, and the populations are closer to $1$ on the left and to $0$ on the right than those of $N=4$. This means that even though the observables are of the same order in both sizes, the larger system favors more the formation of particle domains.

Finally, it is important to note that these results already show that there is no rectification when $\Delta=0$, as the current is $\propto f$. Interactions are thus required in the present scheme for it to behave as a diode.

\subsection{Transport in noninteracting tilted lattice}
Using the exact solution described above, we calculate the transport properties for noninteracting tilted systems larger than those obtained analytically in the previous section. First we consider different values of $f$ and $E$ on lattices of $N=14$, as show in Fig. \ref{delta0_populations}. Similarly to the results for $N=4$ of Eq. \eqref{a_values_N4}, $f$ just modifies the amplitude of the deviation of the populations from the value $1/2$, leaving the overall form of the profile identical. The effect of $E$ (and of $V=EN$, shown in Fig. \ref{fig_domains}(c)) is different, as it changes the form of the profiles. Indeed, as $E$ increases these become more flat on both halves of the lattice, preventing the propagation of particles and thus driving the system to an insulating NESS.

Furthermore, considering Eq. \eqref{relation_a1_b}, the current is proportional to $f$, as in the analytical results of $N=4,5$. Given the latter, we analyze the current for the maximally-driven $f=1$ case only, when varying $V$ and $N$. First we evaluate $\mathcal{J}$ as a function of the system size for several values of $V$; the results are shown in Fig. \ref{delta0_transport}(a). The current decays monotonically both with $V$ and $N$, as expected. Also, for large $V\gtrsim4.5$ an exponential trend with the system size is clear, indicating insulating behavior. For lower $V$ the decay with $N$ is much slower, or even not appreciable in the scale of the plot (as for $V=2$); here we anticipate that the exponential fit will emerge on length scales larger than the system size.

We also study the behavior of the current as a function of $V$ for fixed system sizes; the results are shown in Fig. \ref{delta0_transport}(b). Importantly, for large-enough systems, we find an exponential decay of $\mathcal{J}$ with $V$ from a value that decreases with $N$. These results evidence how rapidly an insulating state develops due to the tilt.

\section{Ansatz for domain NESS} \label{app_ansatz}

\begin{figure*}[t]
\begin{center}
\includegraphics[scale=0.75]{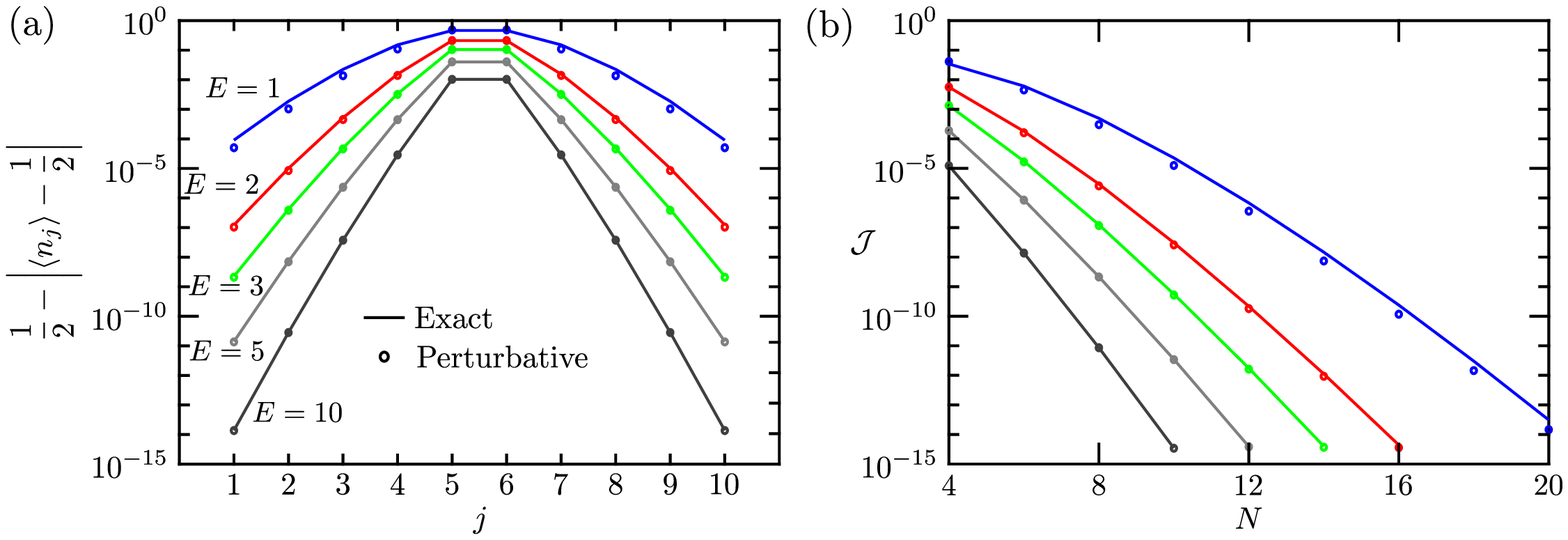}
\end{center}
\caption{Comparison of noninteracting NESS expectation values from exact calculations and insulating Ansatz. (a) Populations across a lattice of $N=10$ and several values of $E$, for exact (solid lines) and perturbative ($\circ$) calculations. (b) Particle currents as a function of the system size; the same convention of colors and symbols of panel (a) is used.}
\label{delta0_ansatz}
\end{figure*}

To unravel the structure of the insulating domain reverse-driving NESSs, consider first $J=0$, where the Hamiltonian eigenstates are products in the Fock basis. The highest energy of the $n$ particle sector corresponds to a single eigenstate with a domain on the right (only degenerate with the domain on the left when $E=0$), denoted as 
$|B_{n}\rangle=|00\ldots0_{N-n}11\ldots1_n\rangle$. Crucially, this state increases its separation in energy from the rest as $E$ or $\Delta$ increase, and is a dark configuration of the $f=-1$ driving \cite{nosotros2013prb}. For low hopping $J\ll \Delta\text{ or }E$ and right-to-left particle transport, the associated eigenstate $|\Psi^D_{n}\rangle$ is perturbatively built from break-away configurations corresponding to the most inner particle moving to the left and the most inner hole to the right. These are denoted as $|\psi_k^n\rangle_{\text{p}}$ and $|\psi_k^n\rangle_{\text{h}}$ respectively for $k$ hopping processes. Due to the energy gap $\Delta+\frac{k}{2}E$ between $|B_{n}\rangle$ and these configurations, the $k$th order correction to $|B_n\rangle$ in perturbation theory is $\sim|\psi_k^n\rangle_{\text{h}}+|\psi_k^n\rangle_{\text{p}}$ with amplitude
\begin{align} \label{coeff_d_with_Delta_tilt}
d_k&=\frac{J^k}{(2\Delta+E)(2\Delta+2E)\cdots(2\Delta+kE)}\notag\\
&=\frac{f_j}{\left(\frac{2\Delta}{E}+1\right)^{(k)}}\left(\frac{J}{E}\right)^{k}.
\end{align}
Here we introduced the Pochhammer symbol $y^{(k)}=y(y+1)(y+2)\cdots(y+k-1)=\Gamma(y+k)/\Gamma(y)$, and a factor $f_j$ different from 1 when the particle (hole) reaches site $j=1$ ($j=N$):
\begin{align}
f_j=\begin{cases}
\frac{2\Delta+(N-n)E}{\Delta+(N-n)E}\quad\text{if}\quad j=1\\
\frac{2\Delta+nE}{\Delta+nE}\quad\text{if}\quad j=N\\
1\qquad\text{otherwise}.
\end{cases}
\end{align}
In the noninteracting ($\Delta=0$) and homogeneous ($E=0$) limits the amplitude reduces to
\begin{align}
d_k=\begin{cases} \label{coeff_d_limits}
\frac{1}{k!}\left(\frac{J}{E}\right)^k=\frac{1}{k!}e^{-k/\xi(E)}\quad\text{if}\quad\Delta=0\\
\left(\frac{J}{2\Delta}\right)^k=e^{-k/\xi(2\Delta)}\quad\text{if}\quad E=0,
\end{cases}
\end{align}
with localization length scale $\xi(Q)=1/\ln(Q/J)$. Thus, in general the corrections to $|B_{n}\rangle$ are suppressed at least exponentially with $k$. In addition, since only the configurations with a particle on site $1$ or a hole on site $N$ are not dark (i.e. they are affected by the $f=-1$ driving), $|\Psi^D_{n}\rangle$ become exponentially close to being dark states of the $f=-1$ driving. Note that an identical description can be performed in the forward $f=1$ driving scheme for the homogeneous interacting case (with degenerate left and right high-energy domains) and the noninteracting tilted system (where the left domain is the state of lowest energy). However, when interactions and tilt coexist this approach is no longer valid.

From the dark states $|\Psi^D_{n}\rangle$ we define the statistical mixture of Eq. \eqref{ness_ansatz}, where the probabilities $p_n$ are determined from a detailed balance condition. Namely, if $P_{n\to m}$ is the transition probability from particle number sector $n$ to sector $m$ due to the driving, the detailed balance condition is $P_{n\to n+1}+P_{n\to n-1}=P_{n-1\to n}+P_{n+1\to n}$, with probabilities
\begin{align}
\begin{split}
&P_{n\to n+1}=p_n\Gamma\langle\Psi_D(n)|\sigma_1^-\sigma_1^+|\Psi_D(n)\rangle\\
&P_{n\to n-1}=p_n\Gamma\langle\Psi_D(n)|\sigma_N^+\sigma_N^-|\Psi_D(n)\rangle.
\end{split}
\end{align}
The solution of these equations for an even system gives
\begin{align} \label{pn_analytic}
p_{\frac{N}{2}+m}=\frac{p}{\prod_{j=-|m|+1}^{|m|-1}\left(\frac{2\Delta}{E}+\frac{N}{2}+j\right)^{2(|m|-|j|)}}\left(\frac{J}{E}\right)^{2|m|^2}
\end{align}
for $-N/2\leq m\leq N/2$. Note that this distribution of probabilities is symmetric, so $p_{\frac{N}{2}+m}=p_{\frac{N}{2}-m}$. Here $p=p_{\frac{N}{2}}$ is fixed by the normalization condition in Eq. \eqref{ness_ansatz}, and is also the highest of all the probabilities $p_n$. Thus, the NESS of the system is largely dominated by the state with occupied and empty domains of equal size.

To provide a benchmark of the ansatz of Eq. \eqref{ness_ansatz}, we reproduced the $f=1$ analytical results of Eq. \eqref{popul_N4_analytic_D0}. In particular, for any even $N\geq4$ and keeping only $m=\pm1$ in Eq. \eqref{pn_analytic}, the probabilities of the ansatz are, up to $\mathcal{O}(E^{-2})$,
\begin{align}
    p_{\frac{N}{2}}=1-8\left(\frac{J}{EN}\right)^2,\qquad p_{\frac{N}{2}\pm1}=4\left(\frac{J}{EN}\right)^2.
\end{align}
This leads to the general expression
\begin{align}
    \left\langle n_{\frac{N}{2}}\right\rangle=1-\left(\frac{J}{E}\right)^2\left[1+\left(\frac{2}{N}\right)^2\right],
\end{align}
which agrees with the result of $\langle n_2\rangle$ of Eq. \eqref{popul_N4_analytic_D0} for $N=4$. We have also reproduced $\langle n_1\rangle$ keeping up to $\mathcal{O}(E^{-4})$. 

\subsection{Accuracy of NESS insulating ansatz}
Now we show that even for small noninteracting tilted systems, the aynsatz of Eq. \eqref{ness_ansatz} captures very well the observed ansulating behavior. For this we compare the values of the deviations of the populations from $1/2$ (Fig. \ref{delta0_ansatz}(a)) and the particle currents (Fig. \ref{delta0_ansatz}(b)) obtained from the ansatz and from the exact solver for lattices of different sizes. Remarkably, even down to $E=1$ the results from the ansatz are very close to the exact ones.  In the presence of interactions, the comparison of both types of calculations has been given in Fig. \ref{fig_recti}(a) for reverse transport, again showing an excellent agreement. These results demonstrate the huge power of the ansatz to describe the nonequilibrium properties of a boundary-driven large-domain insulating NESS.

We emphasize that in spite of the similar description of the domain-like insulating NESSs induced by interactions and tilt on their own, there are some key differences. First, the former requires a minimun value $\Delta=1$, as for lower $\Delta$ the transport is ballistic \cite{Benenti:2009,nosotros2013prb,nosotros2013jstat,landi2021arxiv}; the latter can be achieved with any tilt provided the chain is long enough. Second, as previously discussed, insulating domains emerge for $E=0$ and strong interactions at $f\approx1$ only, while they can do it at any $f$ for $\Delta=0$ and long- or tilted-enough systems.

\section{CP symmetry of tilted interacting model} \label{app_cpsymm}
Hamiltonian \eqref{hami_tilted} with $\mu=-E(N+1)/4$ is invariant under simultaneous charge conjugation and center-reflection operations, e.g. $Un_jU^{\dagger}=1-n_{N-j+1}$ for the CP symmetry operator $U$. Also, since $U^2=I$, its eigenvalues are $\pm1$, corresponding to even and odd symmetry. However, only at half filling $U$ commutes with the total particle number operator $\tilde{N}=\sum_jn_j$, since $U\tilde{N}U^{\dagger}=N-\tilde{N}$. Thus, only in this case we can consider particle number conservation and CP symmetry simultaneously. This property is used to build the Hamiltonian of the even CP half-filled symmetry sector whose energy eigenspectrum features the avoided crossings that coincide with the forward current resonances (see Fig. \ref{fig_reso_4sites}(a)).

Crucially, this does not mean that our giant rectification mechanism relies on the presence of CP symmetry. In fact, the particle transport remains identical when this symmetry is broken by taking $\mu\neq-E(N+1)/4$. This occurs because within the assumed wide-band limit, the energy gap between different particle number sectors (controlled by $\mu$) is irrelevant. Only the energy differences between eigenstates within each sector, which are independent of $\mu$, are important for our device. The CP symmetry just allows us to isolate even and odd eigenstates, transparently evidencing the connection between the avoided crossings and current resonances.\\
In light of this discussion, also note that forward particle transport is identical (up to a direction change) to reverse transport with inverted tilt $-E$.

\subsection{Implementation of CP symmetric Hamiltonian}
To illustrate how the exact eigenstructure of the CP symmetric tilted Hamiltonian is obtained, we take $N=4$. We initially consider the half-filling sector, where the particle number operator $\tilde{N}$ and CP symmetry operator $U$ commute. First we identify each equivalent class (EC), corresponding to each set of basis states that are equivalent under CP symmetry, and their representative state (RS) \cite{jung2020kor}. These are given in table \ref{table1}, along with the period of each EC.\\ 
\begin{table}[ht]
\begin{center}
\begin{tabular}{|c|c|c|}
\hline
 RS & EC & Period \\ 
\hline\hline
 1001 & 1001 & 2 \\
 & 0110 & \\
\hline
 1100 & 1100 & 1 \\ 
\hline
 1010 & 1010 & 1 \\ 
\hline
 0101 & 0101 & 1 \\ 
\hline
 0011 & 0011 & 1 \\ 
\hline
\end{tabular}
\caption{Representative states for $N=4$ half-filled system, states of the corresponding equivalent class and period.}
\label{table1}
\end{center}
\end{table}
The basis elements of the even CP half-filled sector, which are common eigenstates of the number and CP symmetry operator, are thus
\begin{align}
\begin{split}
    &|1001,+\rangle=\frac{1}{\sqrt{2}}(|1001\rangle+|0110\rangle)\\
    &|1100,+\rangle=|1100\rangle,\ \ |1010,+\rangle=|1010\rangle,\\
    &|0101,+\rangle=|0101\rangle,\ \ |0011,+\rangle=|0011\rangle.\\
\end{split}
\end{align}
The Hamiltonian of this symmetry sector, with the given order of basis elements, is
\small
\begin{align}
    H^{2\text{e}}=\frac{1}{4}\begin{pmatrix}
    -\Delta & 0 & 2\sqrt{2}J & 2\sqrt{2}J & 0 \\
    0 & \Delta+4E & 2J & 0 & 0 \\
    2\sqrt{2}J & 2J & -3\Delta+2E & 0 & 0 \\
    2\sqrt{2}J & 0 & 0 & -3\Delta-2E & 2J \\
    0 & 0 & 0 & 2J & \Delta-4E
    \end{pmatrix}.
\end{align}
\normalsize
Its spectrum is depicted in Fig. \ref{fig_reso_4sites}(a); due to the no-crossing theorem \cite{Landau}, it only shows avoided crossings. On the other hand, the half-filled odd CP symmetric sector only has one basis element, namely
\begin{align}
    |1001,-\rangle=\frac{1}{\sqrt{2}}(|1001\rangle-|0110\rangle),
\end{align}
with energy $-\Delta/4$. This completes the 6 states of the half-filling sector.

Beyond half filling the CP symmetry operator does not commute with $\tilde{N}$. Thus we can build sectors of fixed number of particles or fixed symmetry, but not both. However, each case has redundant information. In the former scenario, the Hamiltonian of $m$ particles is equal to that of $m$ holes. In the latter, the Hamiltonians of even and odd CP symmetric sectors are also identical. To see this, note that the basis elements of each sector are the $\pm$ equal superpositions of the states of each EC listed in table \ref{table2}, namely 
\begin{table}[ht]
\begin{center}
\begin{tabular}{|c|c|c|}
\hline
 RS & EC & Period \\ 
\hline\hline
 1111 & 1111 & 2 \\
 & 0000 & \\
\hline
 1110 & 1110 & 2 \\ 
 & 1000 & \\
\hline
 1101 & 1101 & 2 \\ 
 & 0100 & \\
\hline
 1011 & 1011 & 2 \\ 
 & 0010 & \\
\hline
 0111 & 0111 & 2 \\ 
 & 0001 & \\
\hline
\end{tabular}
\caption{Representative states for $N=4$ beyond half filling, states of the corresponding equivalent class and period.}
\label{table2}
\end{center}
\end{table}
\begin{align}
\begin{split}
|1111,\pm\rangle=\frac{1}{\sqrt{2}}(|1111\rangle\pm|0000\rangle)\\
|1110,\pm\rangle=\frac{1}{\sqrt{2}}(|1110\rangle\pm|1000\rangle)\\
|1101,\pm\rangle=\frac{1}{\sqrt{2}}(|1101\rangle\pm|0100\rangle)\\
|1011,\pm\rangle=\frac{1}{\sqrt{2}}(|1011\rangle\pm|0010\rangle)\\
|0111,\pm\rangle=\frac{1}{\sqrt{2}}(|0111\rangle\pm|0001\rangle).
\end{split}
\end{align}
The $\pm$ superposition states of $0$ and $4$ particles are disconnected from the other basis elements, and have energy $3\Delta/4$. The states of 1 and 3 particles are connected with each other by the Hamiltonian, which has the matrix representation
\begin{align}
    H^{\text{e/o}}=\frac{1}{4}\begin{pmatrix}
    \Delta+3E & 2J & 0 & 0 \\
    2J & -\Delta+E & 2J & 0 \\
    0 & 2J & -\Delta-E & 2J \\
    0 & 0 & 2J & \Delta-3E
    \end{pmatrix}.
\end{align}
Similarly to the half-filled case, the spectrum of this Hamiltonian features avoided crossings only. This completes the 10 states of the sectors beyond half filling, and thus the 16 states of the $N=4$ system.\\
The above construction can be directly extended to larger systems. This is exemplified in Fig. \ref{fig_spectrum_10sites} for the even CP symmetry sector of 10 sites, which corresponds to results presented in Fig. \ref{fig_recti}(d)-(f).

\section{Perturbative calculation of eigenenergies} \label{app_pert_eigen}

\begin{figure}[t]
\begin{center}
\includegraphics[scale=0.7]{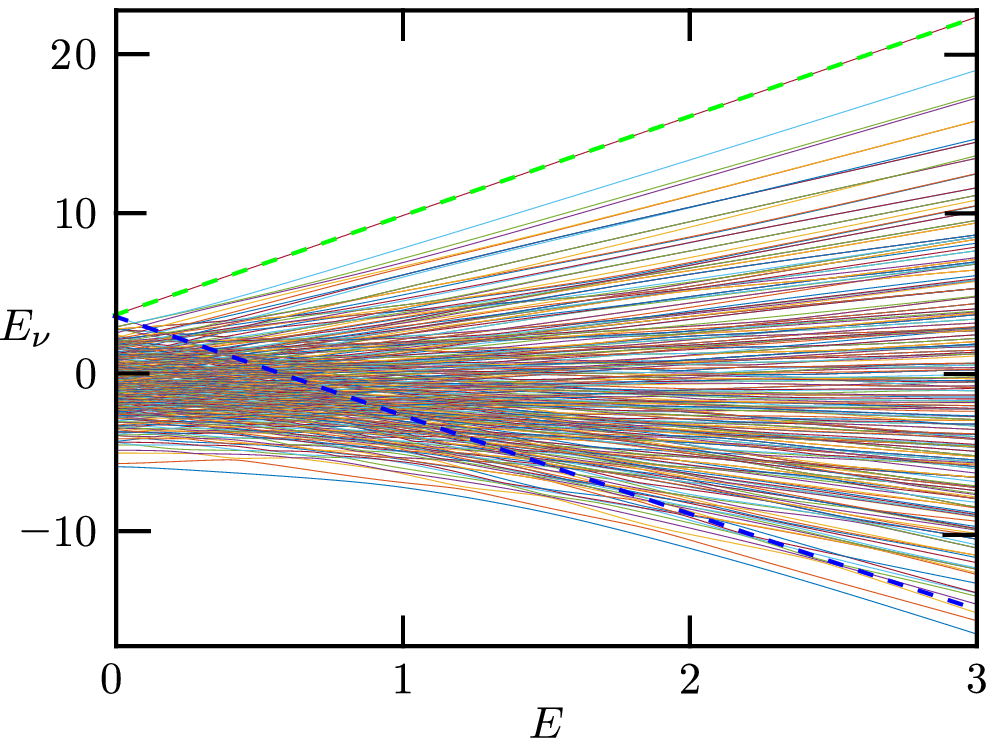}
\end{center}
\caption{Eigenstructure of half-filled even CP symmetric sector for $N=10$ and $\Delta=2$. The solid lines correspond to the eigenenergies, and the dashed lines to the second-order perturbative corrections to the domains states pinned to the right (green) and left (blue) boundaries, namely Eqs. \eqref{domain_perturb_energies}.}
\label{fig_spectrum_10sites}
\end{figure}

Given the many-body nature of our problem, it is not feasible to obtain exactly the eigenstructure of the Hamiltonian for large systems, even with the explicit use of the CP symmetry described in the previous subsection. Considering that our giant rectification mechanism manifests for strong interactions, we perform a perturbative calculation of the most important eigenstates in this limit. For this we consider first the case of zero hopping, where the eigenstates are product states in the Fock basis. Here, the energies for $N$ sites and domains of $n$ particles pinned to the left (L) and right (R) boundaries are
\begin{align}
\begin{split}
    E_{n,\text{L}}=\frac{1}{4}\left[\Delta(N-3)-En(N-n)\right],\\
    E_{n,\text{R}}=\frac{1}{4}\left[\Delta(N-3)+En(N-n)\right].
\end{split}    
\end{align}
Clearly both states are degenerate and have the highest energy of the $n$ particle sector in the absence of tilt \cite{nosotros2013prb}. For a finite tilt, $E_{n,\text{R}}$ is the highest eigenstate of the $n$ particle sector, while $E_{n,\text{L}}$ is the second highest one for low $E$ and the lowest one for very large $E$.\\
When turning on a very weak hopping $J$, these energies are corrected by second-order processes of the most inner particle jumping to the neighboring empty site and then jumping back. Assuming no degenerate states, the corrected energies are
\begin{align} \label{domain_perturb_energies}
\begin{split}
    \tilde{E}_{n,\text{L}}=E_{n,\text{L}}+\frac{J^2}{4\Delta-2E},\\
    \tilde{E}_{n,\text{R}}=E_{n,\text{R}}+\frac{J^2}{4\Delta+2E}.        
\end{split}
\end{align}
For $N=4$, we identify $\tilde{E}_{2,\text{R}}=\tilde{E}_{0011}$ as the energy $\nu=5$ of the half-filled even CP eigenstructure of Fig. \ref{fig_reso_4sites}(a), which monotonically increases with $E$. Also, far from $E=2\Delta$, $\tilde{E}_{2,\text{L}}=\tilde{E}_{1100}$ corresponds to the eigenenergy whose avoided crossings in the exact spectrum match the forward transport resonances, namely to $\nu=4$ for $0\leq E\lesssim2$, $\nu=3$ for $2\lesssim E\lesssim4$, $\nu=2$ for $4\lesssim E\lesssim8$ and $\nu=1$ for $E\gg1$. A similar description holds for larger systems; this is seen in Fig. \ref{fig_spectrum_10sites}, where we plot $\tilde{E}_{5,\text{R}}$ and $\tilde{E}_{5,\text{L}}$ for $N=10$ sites, indicated by green and blue dashed lines respectively.\\ 
We also focus on the case of $L=4$ to perturbatively calculate the other energies associated to the basis elements of the half-filled even CP symmetry sector. In the zero hopping limit, these are
\begin{align}
\begin{split}
    &E_{1010}=E_{\nu=1}=-\frac{1}{4}(2E+3\Delta),\\
    &E_{0101}=E_{\nu=2}=\frac{1}{4}(2E-3\Delta),\\
    &E_{1001,+}=E_{\nu=3}=-\frac{\Delta}{4}.
\end{split}    
\end{align}
Turning on a weak hopping, and keeping up to $\mathcal{O}(J^2)$, the corrected energies are
\begin{align}
\begin{split}
    &\tilde{E}_{1010}=E_{1010}+\frac{J^2}{2}\left[\frac{1}{E-2\Delta}-\frac{2}{E+\Delta}\right],\\
    &\tilde{E}_{0101}=E_{0101}+\frac{J^2}{2}\left[\frac{2}{E-\Delta}-\frac{1}{E+2\Delta}\right],\\
    &\tilde{E}_{1001,+}=E_{1001,+}+J^2\left[\frac{1}{\Delta+E}+\frac{1}{\Delta-E}\right].
\end{split}    
\end{align}
For tilts lower than those of the divergences ($E=\Delta,2\Delta$), these expressions work reasonably well, as seen in Fig. \ref{fig_reso_4sites}(a). In fact, the first crossing of $\tilde{E}_{1100}$ and $\tilde{E}_{1001,+}$ takes place at $E\approx2.08$, which is very close to the location of the first forward current resonance, $E=2.15$.

\section{Eigenstate population profiles} \label{app_eigen_popul}

\begin{figure}[t]
\begin{center}
\includegraphics[scale=0.6]{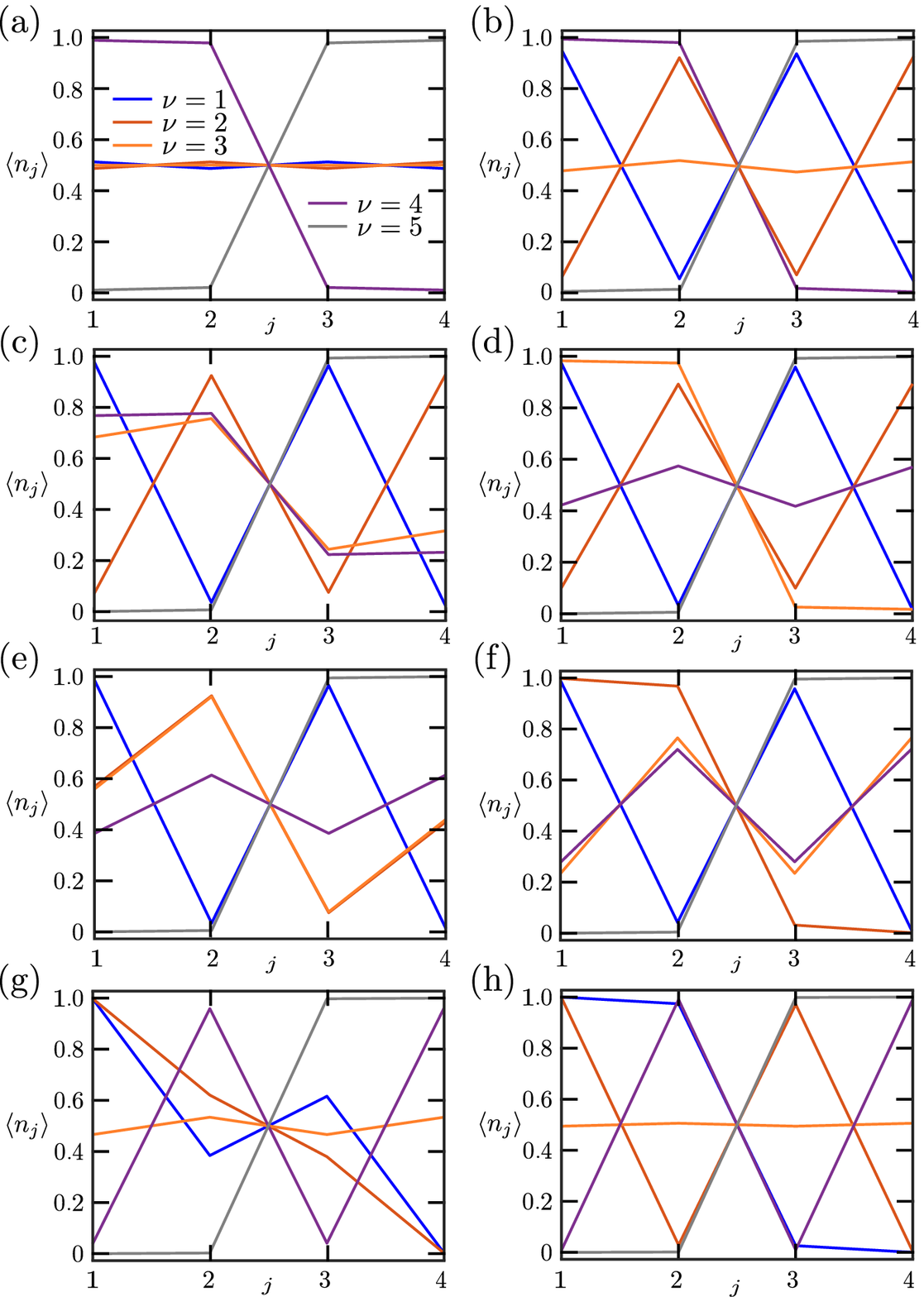}
\end{center}
\caption{Site population $\langle n_j\rangle$ of the Hamiltonian eigenstates of the even CP symmetry half-filled sector for a chain of $N=4$, $\Delta=5$, and several values of tilt. (a) $E=0.01$. (b) $E=1$. (c) $E=2.15$. (d) $E=3$. (e) $E=3.67$. (f) $E=5$. (g) $E=9.65$. (h) $E=16$.}
\label{fig_eigenvec_popul}
\end{figure}

To further illustrate the mechanism underlying the forward transport resonances, we show in Fig. \ref{fig_eigenvec_popul}, for several values of the tilt, the population of the five eigenstates of the even CP symmetry half-filled sector discussed on Fig. \ref{fig_reso_4sites}. For a small tilt (Figs. \ref{fig_eigenvec_popul}(a) and (b)), eigenstates $\nu=4,5$ feature domain profiles pinned to the left and right boundaries respectively; the latter is always present and becomes more polarized as $E$ increases, while the former is strongly modified by $E$. Namely, as the first resonance is approached (Fig. \ref{fig_eigenvec_popul}(c)), the left domain of $\nu=4$ is degraded and the population of $\nu=3$ starts developing a domain profile, which is fully established after crossing the resonance (Fig. \ref{fig_eigenvec_popul}(d)). A similar pattern of domain suppression and population exchange among the eigenstates involved in the corresponding avoided crossing takes place at the second (Figs. \ref{fig_eigenvec_popul}(e) and (f)) and third (Figs. \ref{fig_eigenvec_popul}(g) and (h)) resonances. Finally, at very large tilt it is $\nu=1$, the associated eigenstate of highest population, the one that features a left domain (Fig. \ref{fig_eigenvec_popul}(h)). Clearly, the absence of the left domain at the resonant values of $E$ for the eigenstates $\nu$ of highest probability at the NESS (see Fig. \ref{fig_reso_4sites}(b)) is in line with our picture for forward transport.

\section{Rectification beyond Lindblad driving: Mesoscopic leads approach} \label{app_mesoreservoirs}

\begin{figure*}[t]
\begin{center}
\includegraphics[scale=0.9]{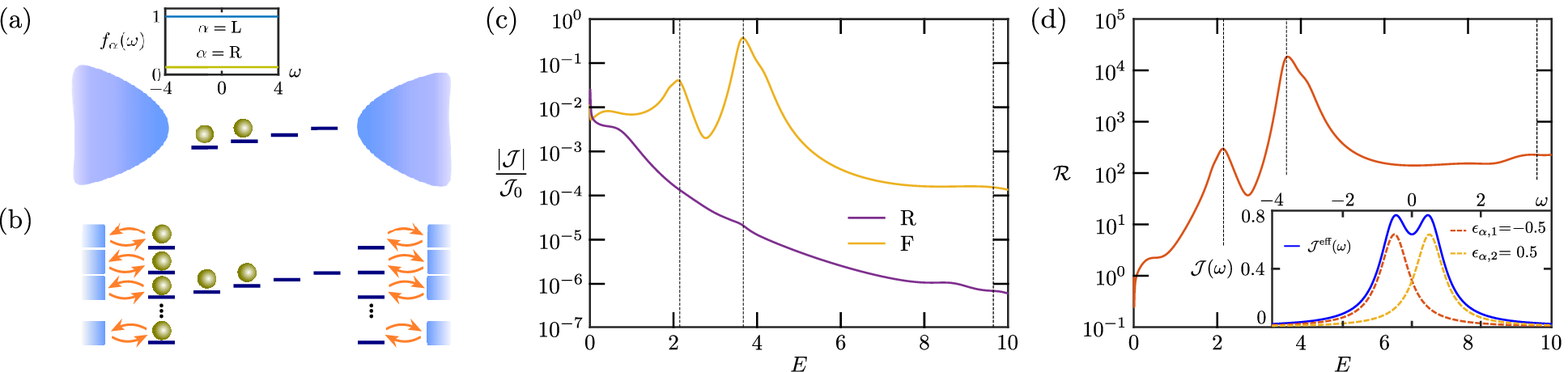}
\end{center}
\caption{Rectification with mesoscopic leads approach. (a) Scheme of tilted system coupled to infinite fermionic reservoirs at its boundaries. Inset: Fermi-Dirac distributions of left (L) and right (R) reservoirs for completely occupied and empty cases respectively. (b) Mesoscopic leads description, in which the infinite reservoirs are substituted by a finite collection of fermionic lead modes, each one connected to its own Markovian bath (blue squares). (c) Forward (F) and reverse (R) particle currents in a lattice of $N = 4$ and $\Delta=5$ as a function of the tilt. We considered $L=2$ lead modes per reservoir, with parameters $T_{\text{L,R}}=10$, $\mu_{\text{L}}=-\mu_{\text{R}}=100$, $\epsilon_{\text{L/R},k}=\pm0.5$ and $\gamma_{\text{L/R},k}=1$. Currents have been rescaled by the maximum current $\mathcal{J}_0$ obtained with the same driving (namely, equal parameters of the lead modes) for a noninteracting homogeneous system. The vertical dashed lines indicate the location of the maximal currents for the Lindblad driving scenario. (d) Corresponding rectification coefficient $\mathcal{R}$. Inset: Effective spectral density \eqref{eff_spectral} of the discretized leads, with the contributions from each mode.}
\label{fig_mesoreservoirs}
\end{figure*}

Here we show that the rectification mechanism introduced in this work also emerges in more realistic nonequilibrium setups than that of Lindblad boundary driving. The latter corresponds to an idealized limit of infinite high-temperature reservoirs with flat and very wide spectral density \cite{Benenti:2009}. Now, we move away from it by employing the so-called \textit{mesoscopic leads} description. This approach decomposes infinite fermionic environments at given temperatures and chemical potentials in terms of a finite number of lead modes, each of which is coupled to a residual bath described by Lindblad damping \cite{nosotros2020prx,marlon2023prb,de2022arxiv}. In brief, we consider the tilted system of interacting spinless fermions, with the Hamiltonian $H$ of Eq. \eqref{hami_tilted}, coupled at its boundaries to macroscopic reservoirs of noninteracting fermions at temperatures $T_{\alpha}=1/\beta_{\alpha}$ and chemical potentials $\mu_{\alpha}$ ($\hbar=k_{\text{B}}=1$, $\alpha=\text{L,R}$). This setup is depicted in Fig. \ref{fig_mesoreservoirs}(a), and described by the total Hamiltonian
\begin{align}
H_{\text{tot}}=H+\sum_{\alpha}\left(H_{\alpha}+H_{S\alpha}\right).
\end{align}
Here, $H_{\alpha}$ is the Hamiltonian of reservoir $\alpha$, and $H_{S\alpha}$ corresponds to its coupling to the tilted system; the latter is modeled as hopping processes from each reservoir mode to the closest site of the lattice. Each reservoir is characterized by its Fermi-Dirac distribution $f_{\alpha}(\omega)=\left(e^{\beta_{\alpha}(\omega-\mu_{\alpha})}+1\right)^{-1}$ and its spectral density
\begin{align} \label{spectral_original}
\mathcal{J}_{\alpha}(\omega)=2\pi\sum_{m=1}^{\infty}|\lambda_{\alpha,m}|^2\delta(\omega-\omega_{\alpha,m}),
\end{align}
with $\omega_{\alpha,m}$ the energy of mode $m$ of reservoir $\alpha$, and $\lambda_{\alpha,m}$ its coupling to the closest site of the lattice. 
Obtaining the dynamics and steady state of this boundary-driven system is extremely challenging. Interactions between its particles generally make the problem intractable from analytical methods; in addition, finite temperatures and strong coupling to the reservoirs can induce large non-Markovian effects.

The mesoscopic leads approach has been recently shown to successfully overcome these complexities \cite{nosotros2020prx}. Its key idea consists of reducing the macroscopic reservoirs to a finite collection of $L$ lead modes, each one connected to its own Markovian bath, as depicted in Fig. \ref{fig_mesoreservoirs}(b). In this scenario, each mode contributes to the effective spectral density with a Lorentzian function:
\begin{align} \label{eff_spectral}
\mathcal{J}^{\text{eff}}_{\alpha}(\omega)=\sum_{k=1}^L\frac{|\kappa_{\alpha,k}|^2\gamma_{\alpha,k}}{(\omega-\epsilon_{\alpha,k})^2+(\omega_{\alpha,k}/2)^2}.
\end{align}
Here, $\epsilon_{\alpha,k}$ is the energy of mode $k$ of lead $\alpha$, $\kappa_{\alpha,k}$ its coupling to the closest site of the lattice, and $\gamma_{\alpha,k}$ its Markovian dissipation rate. By carefully selecting the number of lead modes $L$ and their corresponding parameters \cite{nosotros2020prx,de2022arxiv}, the original spectral density characterizing the system-reservoir couplings (Eq. \eqref{spectral_original}) can be accurately approximated from Eq. \eqref{eff_spectral}.

For simplicity, we calculate the NESS of an interacting tilted lattice of $N=4$ sites, using a minimal but nontrivial realization of this updated driving setup with $L=2$ modes per lead. This way, the NESS can be obtained with exact diagonalization. To mimic the conditions of Fig. \ref{setup}(d), we consider $\Delta=5$, high temperatures $T_{\text{L,R}}=10$, chemical potentials $\mu_{\text{L}}=-\mu_{\text{R}}=100$ to induce large bias for forward transport (so the left (right) bath is almost completely full (empty), see inset of Fig. \ref{fig_mesoreservoirs}(a)), and $\mu_{\text{L,R}}$ with opposite signs for reverse transport. For both reservoirs we consider lead mode energies $\epsilon_{\text{L/R},k}=\pm0.5$, and dissipation rates $\gamma_{\text{L/R},k}=1$. The effective spectral density is the sum of two Lorentzians depicted in the inset of Fig. \ref{fig_mesoreservoirs}(d). The currents and rectification coefficient are shown in Figs. \ref{fig_mesoreservoirs}(c) and (d) respectively.

Just as in the Lindblad driving scenario, an increasing tilt impedes reverse transport and leads to large resonances of the forward current. Furthermore, the latter agree with the avoided crossings of the eigenspectrum. Thus, the giant rectification mechanism reported in this work is not an artefact of the Lindblad driving, but generally emerges from the interplay between strong particle interactions, tilt and large bias. A full study of charge and thermal quantum diodes in this more realistic setup, considering larger leads and different driving conditions, will be performed in a future work.

\bibliographystyle{unsrt}

\bibliography{rectification_bib}

\end{document}